\documentclass
    [a4paper,10pt]{article}
\usepackage{graphicx}
\usepackage{amssymb}
\usepackage[utf8]{inputenc}
\usepackage[english]{babel}
\usepackage{xcolor}

\title{Atomic Size Pearls being Dark Matter giving Electron Signal }
\author{H.B. Nielsen,
  Niels Bohr Institut,
C.D.Froggatt, Glasgow University}
\date{   2021}

\title{Atomic Size Pearls being Dark Matter and giving Electron Signal}
\author{C. D. Froggatt, Glasgow University\\
H.B.Nielsen, Niels Bohr Institute}

\begin{document}

\maketitle
\begin{center}
  Extended version of the \\
  Contribution to Proceedings to the 24th Workshop
  ``What Comes Beyond the Standard Models'', Bled, July 3.-- 11., 2021\\
\end{center}
\begin{abstract}
We seek to explain both the seeming
  observation of dark matter by the seasonal variation
   of the DAMA-LIBRA data and the observation of
  ``electron recoil'' events at Xenon1T in which the liquid-Xe-scintillator was
  excited by electrons - in excess to the expected background - by the
  {\em same} dark matter model.
   In our model the dark matter consists
   of bubbles of a new type of vacuum containing ordinary atomic matter,
   say diamond,
  under high pressure ensured by
  the surface tension of the separation  surface (domain wall).
  This atomic matter is surrounded by a cloud of electrons extending
  out to about atomic size.
  We also seek to explain the self interactions of
  dark matter suggested by astronomical studies of dwarf galaxies and the
  central structure of
  galaxy clusters. At the same time we consider the interaction with matter
  in the shielding
  responsible for slowing the dark matter down to a low terminal velocity,
  so that collisions with nuclei
  in the underground detectors have insufficient energy to be detected.
  Further we explain the ``mysterious'' X-ray line of 3.5 keV from
  our dark matter particles colliding with each other so that the
  surfaces/skins
  unite. Even the 3.5 keV X-ray radiation from the Tycho supernova remnant
  is explained as
  our pearls hitting cosmic rays in the remnant.

  What the DAMA-LIBRA and Xenon1T experiments see is supposed to be our dark
  matter pearls
  excited during their stopping in the shielding or the air. The most
  remarkable support for our type of model is that both these underground
  experiments see events with about 3.5 keV energy, just the energy of the
  X-ray line.


  We fit numerically the cross section over mass ratio for the self interaction
  of the dark matter at low velocity $v \rightarrow 0$, as observed in the study of dwarf galaxies and find the
  size of the pearls to be rather  close to the smallest possible under the restrictions
  from our earlier 3.5 keV line fit.
  However the mass obtained from this dwarf galaxy fit turns out to be too small
  for allowing the pearls to penetrate down to the DAMA experiment in less than
  a year, as needed to avoid a washing out of the seasonal variation.
  This led us to investigate more carefully the velocity relevant to our calculation and to consider pearls of larger size.
  
  Also the total energy of the dark matter pearls stopped in the shield is
  reasonably matching order of magnitudewise with the absolute observation
  rates of DAMA-LIBRA and Xenon1T, although the proposed explanation of
  their ratio 
  requires further development.

  It should be stressed that accepting that the different phases of the vacuum
  could be realized inside the Standard Model, our whole scheme could be
  realized inside the Standard Model. So then no new physics is needed for dark
  matter!

\end{abstract}

\section{Introduction}

For a long time we have worked on a dark matter model
\cite{Dark1, Dark2, Tunguska, supernova, Corfu2017, Corfu2019, theline,
  Bled20}, in which the dark
matter consisted of cm-size pearls which were in fact bubbles of a  new
vacuum type surrounded by a skin caused by the surface tension of this
new vacuum. This skin kept a piece of usual atomic matter highly
compressed inside the bubble.
The present article is an extension and update of our Bled-proceedings article
2021 \cite{Bled21} 
by improving the calculation of the radius of the pearl and of the surrounding
electron cloud.
Further we have here a discussion of the
troubles in our model of getting the dark matter pearls, with the 
rather large interaction suggested by studies of dwarf galaxies, through the 1400 m shielding
of the DAMA experiment in less than a year.

The idea of our model is actually rather similar to the 
Dark Matter as Color Superconductors proposed by Ariel Zhitnitsky
\cite{Zhitnitsky} in 2003. This was called to our attention by
Konstantin Zioutas with special interest in the solar corona heating
\cite{solar}.

In fitting data in the old model - in which the pearls were supposed so
large and heavy that they could have caused the Tunguska event -
the most and almost only successful fit consisted in that we
fitted, with a common parameter, both the overall rate and the very 3.5 keV
energy of the X ray line
originally observed in several galaxy clusters, Andromeda and the
Milky Way Center \cite{Bulbul, Boyarsky, Boyarsky2, Bhargava, Sicilian, Foster}
and supposedly coming from dark matter.
But now it turned out that this successful fitting relation between the
3.5 keV energy and the overall rate of the X-ray radiation only depends on
the density of the pearls or equivalently the Fermi momentum
or energy of the electrons kept inside the pearls, but not on the absolute
size of the pearls. Thus we could change the model to make the pearl sizes
much smaller,
and that is what we did in our contribution to the Bled workshop proceedings
in 2021.
So the pearls making up the dark matter are
now rather of
atomic size. Really we shall consider pearls with radii ranging from
$R \sim 10^{-11}m$ to $R \sim 10^{-9}m$.
But even such small pearls get stopped
to some extent by the shielding into which they must penetrate to reach
the underground experiments like the DAMA-LIBRA and Xenon experiments looking
for dark matter. Using an astronomical observation based model by
Correa \cite{CAC} especially, we shall construct a rather definite picture
of our pearls from which we estimate that the pearls hitting the
earth actually get stopped presumably in the atmosphere, but if not there then
at least in the earth shielding. The pearls thereby lose so much speed that
it becomes
quite understandable that the Xenon-experiments, looking for nuclei
being hit by them and causing scintillation in fluid xenon, will not see
any such events. However the DAMA-LIBRA experiment \cite{DAMA1, DAMA2} would not
distinguish if it is a nucleus that is hit or some energy is released which
causes the scintillator to luminesce. So only the DAMA-LIBRA experiment
would be able to
get a signal if the dark matter, e.g. our pearls, could be somehow
excited and emit their excitation energy when they pass through the detector.
In our model we shall indeed suggest that the pearls get excited
and emit their energy by electron emission. That would not be
easy to distinguish for DAMA-LIBRA but would still of course come with
seasonal variation\footnote{We note however that the ANAIS experiment
has failed to see an annual modulation with NAI(Tl) scintillators and their
results \cite{ANAIS} are incompatible with the DAMA-LIBRA results at
$3.3\sigma$.}
so that it would be observed as dark matter by
DAMA-LIBRA. Whether the emission is via electrons or nuclei would not matter.
But for the Xenon-experiments such electron emission was effectively
not counted for a long time, but now rather recently the Xenon1T experiment has
actually observed an excess of ``electron recoil events''. So they
have now in fact seen an electron emission somehow.

We shall see in section \ref{sec7} that both  the excess
of electron recoil events in Xenon1T \cite{Xenon1Texcess} and the events seen by
DAMA-LIBRA \cite{DAMA1, DAMA2} have the energy of each event remarkably
enough centering about
the energy value 3.5 keV of the mysterious X-ray line found astronomically!

This coincidence of course strongly suggests that these events
from DAMA-LIBRA and Xenon1T are related to
dark matter particles that can be excited precisely by this energy
3.5 keV.

In our earlier papers \cite{Corfu2017, Corfu2019, theline} we have already
connected the excitability of our pearls by
just this energy 3.5 keV and especially the emission of photons
(or here in the present work also electrons) with just this energy with
a gap in the single particle electron spectrum of the pearls
caused by what we call the homolumo-gap effect.

   

A very serious warning, which needs an explanation in order to rescue our model,
is delivered by the fact that if as we now suggest the Xenon1T electron recoil event excess
is coming from just the same decay of dark matter excitations as the DAMA-LIBRA
observation, then these two experiments ought a priori to see equally many
events, say per kg. However, DAMA-LIBRA sees 250 times as many events as
Xenon1T sees excess events.

We shall discuss in section \ref{discuss}  the difficulty due to
this ratio not being unity, but the hope
for now is that the Xenon1T experiment has the observed decaying pearls
falling through a fluid, namely the fluid xenon, while the scintillator in
DAMA-LIBRA is a solid made from NaI(Tl). The pearls are likely to
form a little Xe-fluid bubble around them and flow or fall through the
xenon-fluid, while they will much more easily get caught so as to almost
sit still
or only move much slower through the NaI scintillator. If so the pearls
with their supposed excitations would spend much more time in the
DAMA-LIBRA NaI than in a corresponding volume of xenon-liquid.
Then, hoping for such a difference between the fluid xenon and the solid
NaI, the terminal velocity in the gravitational  field in the
xenon has to be 250 times faster than that in the solid in which it anyway
has to be fast enough to reach through in less than a year. Crudely estimating
the viscosity of fluid xenon then gives limits for the mass of the pearl.

In the following section \ref{Pearl} we describe how, in the Bled 
Proceedings paper \cite{Bled21}, we imagined the particles making up
the dark matter in our model to be bubbles of the size $R=
r_{cloud \; 3.3MeV} \approx 8*10^{-12}m$
with heavy atomic matter inside. These bubbles  
are surrounded by a cloud of electrons, which we show in section \ref{new} 
to have a thickness of at least 
$7.8*10^{-13}m$. 
But presumably the cloud of electrons outside is rather
similar to an ordinary atom, having a thickness of the order of an atomic radius
$ \sim 10^{-10}m$.
In \cite{Bled21} we supposed that the homolumo gap meant that the 
electrons only extended
out to where the electric potential would equal the value 3.5 keV
(the corresponding radius is called $r_{cloud \; 3.5 keV}$).

Here the quantities 3.3 MeV and
3.5 keV in the subscripts are the
numerical electric potentials felt by an electron at the distances mentioned.
A special point to note in this section already present in the earlier articles
about the big pearls is the above mentioned homolumo-gap effect, 
causing a band or gap in
the energy
levels without any single particle electron eigenstates.  The width of this
gap is fitted to
the 3.5 keV line in the observed X-ray spectrum from galaxy clusters, the
Milky Way Center etc. \cite{Bulbul, Boyarsky, Boyarsky2}.

Next in section \ref{ngi} we briefly review astronomical
observations and modelling of the dark matter, which suggests
the idea that dark matter interacts with itself (self interacting
dark matter SIDM).
It is only when the corresponding cross section $\sigma$
is divided by the
particle mass $M$ that we have a combination that has any chance of 
being observed
by its effects on the atomic matter. 

New material in this paper compared to the Bled proceedings \cite{Bled21}
demonstrates that there is a minimal size for a pearl containing electrons 
with a Fermi momentum $p_f = 3.3$ MeV as determined from the intensity and energy of the 3.5 keV X-ray line emitted by galactic clusters
\cite{Corfu2019, theline, Bled20}.
This is presented in section \ref{new}, in which
  we see that indeed the actual pearl size found in \cite{Bled21} is only a little bigger than the minimal size. 

  In section \ref{range} we discuss the range out to which the electron cloud
  may go, a question very important for the cross section $\sigma$ for the
  self interaction and also for the stopping of the pearls once hitting the
  earth. We calculate the radii $r_{cloud \; 3.5 keV}$ and $r_{cloud \; 3.3 MeV}$
  using the same Thomas-Fermi approximation as in \cite{Bled21}. We improve 
  this approximation in section \ref{correction}.
 
  
In  section \ref{Achievements} we list a series of numerical successes of our
model for the dark matter, hopefully making the reader see that there is really
some reason for it being at least in some respects correct.

In section \ref{Impact} we stress again that our dark matter pearls get stopped and
at the same time excited, mainly to emit quanta of energy 3.5 keV, in the air
and/or in the shielding above the experiments. 
It is the braking energy from this
slowing down that is supposed to feed the excitations.

In section \ref{discuss} we investigate the problem for our model that it
must at least get the pearls to reach down through the earth to DAMA in preferably less
than a year, since otherwise one would not observe much seasonal variation.
But for the pearls to get down fast enough we require pearls of a mass too 
high to match well with the amount of self interaction observed at low velocity
$v \to 0$
in the study of dwarf galaxies \cite{CAC}. We are thereby led to consider
pearls of size $R \sim 1 \, nm$ and to speculate that during their passage
through space they may have collected some dirt around them.

A special estimation, based on energy considerations, of whether the number
of events seen by
DAMA-LIBRA and by the Xenon1T electron recoil excess are of a
reasonable order of magnitude is put forward in section \ref{sec5}.
The success of such an estimation has to be rather limited in as far as the
rates of the two
observations - that should have been the same if we do not include the
possibility of
faster or slower motion through the detectors - deviate by a factor of 250.

In section \ref{sec7} we call attention to the perhaps most remarkable
fact supporting a major aspect of our model: That the energy per event
for both DAMA-LIBRA and the Xenon1T-electron recoil excess centers around
3.5 keV, just the energy of the photons in the mysterious X-ray line
seen in galactic clusters mentioned above! So all three effects should
correspond to the emission of an electron or photon due to the same energy
transition inside dark matter.

The values of the parameters characterising our pearls are discussed in 
section \ref{parameters}.
Finally in section \ref{Conclusion} we conclude and provide a short outlook.

\section{Pearl}
\label{Pearl}

  {\bf Dark Matter Atomic Size Pearls, Electronic 3.5 keV Signal }

\begin{figure}
\includegraphics[scale=0.9]{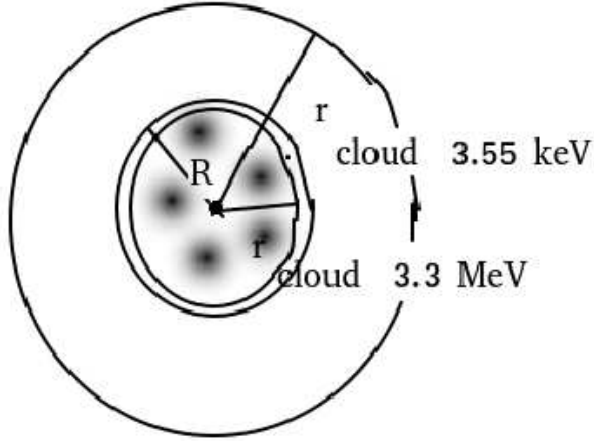}
\caption  {\label{Pearl1} The figure illustrates the bit smaller than atom-size
    complicated/macroscopic dark matter particle in our
    model, a pearl.}
\end{figure}

We here consider minimal size pearls similar to those discussed in Bled \cite{Bled21}; the structure of such small dark matter pearls is sketched
in Figure \ref{Pearl1}. We will later in the article consider larger 
pearls with a radius $R \sim 1 \, nm$.
  \begin{itemize}
  \item In the middle is a spherical bubble of radius
    \begin{eqnarray}
      R \approx r_{cloud \; 3.3 MeV}&\approx & 8*10^{-12}m.
    \end{eqnarray}
    Here $r_{cloud \; 3.3 MeV}$ denotes the radius where the electron potential
    is 3.3 MeV, which is identified with the Fermi energy $E_f$ of the
    electrons in the bulk of the pearl - i.e. inside the radius $R$. We
    estimated the value $E_f = 3.3$ MeV in previous papers \cite{Corfu2019, theline, Bled20} by fitting the
    overall rate of the intensity of the 3.5 keV line emitted by galactic
    clusters and the very frequency 3.5 keV of the radiation in our model.
  \item The outer radius
  \begin{equation}
  r_{cloud \; 3.5 keV} \approx 7*10^{-11} m
  \end{equation}
  is where the electron potential
    is 3.5 keV.  By our story of the ``homolumo gap'':  the
    electron density crudely goes to zero at this radius. (It gradually falls
    in the range between $r_{cloud \; 3.3 MeV} $ and
    $r_{cloud \; 3.5 keV}$).
    \end{itemize}

  {\bf The electron density and potential in the pearls}

\begin{figure}
	  \includegraphics[scale=0.8]{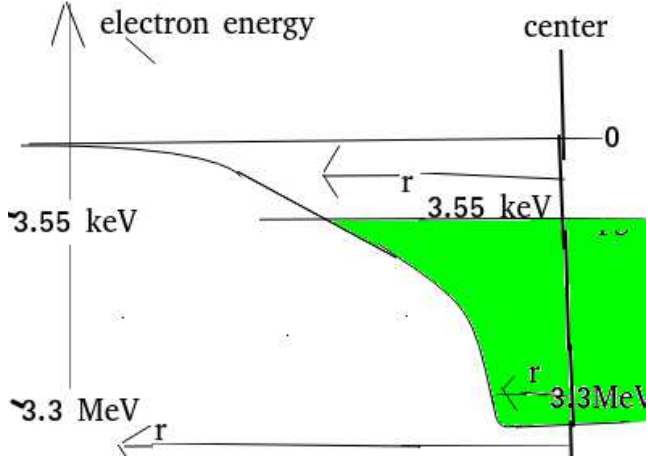}
          \caption{\label{density} Explaining the electron density and electric potential in the Pearl}
\end{figure}

  \begin{itemize}
  \item Due to an effect, we call the {\em homolumo-gap effect} \cite{Corfu2017, jahn}, the nuclei
    in the
    bubble region and the electrons themselves become arranged in such a way
    as to prevent there from being any levels in an interval of width
    3.5 keV. So, as illustrated in Figure \ref{density},
    outside the distance $r_{3.5 keV} = r_{cloud \; 3.5 keV}$ from
    the center of the pearl at which the Coulomb potential is $\sim$ 3.5 keV
    deep there are essentially ($\sim$  in the Thomas-Fermi approximation) no
    more electrons in the pearl-object.
  \item The radius $r_{3.3 MeV}= r_{cloud \; 3.3 MeV}$ at which the potential
    felt by an electron is 3.3 MeV deep, is  supposed to be just the radius
    to which the many nuclei inside the pearl (which replace
    the single nucleus in ordinary atoms) reach out. So inside the bubble
    the potential is much more flat.
  \end{itemize}

  \begin{itemize}
  \item The energy difference between the zero energy line and the effective
    Fermi surface, above which there are no more electrons, is of order
    3.5 keV, the energy so crucial in our work.

  \item Since in the Thomas-Fermi approximation with a homolumo gap 
  there are no electrons outside
    roughly the
    radius $r_{3.5 keV}=r_{cloud \; 3.5 keV}$, this radius will give the maximal
    cross section, even for very low velocity $\sigma_{v\rightarrow 0}$. 
    However we now believe that there will be a tail of the electron cloud further out responsible for the low velocity cross section or even that
    the pearl collects some dirt around it. 
    See section \ref{discuss}. 
    \end{itemize}

  {\bf The homolumo gap effect.}

\begin{figure}
    \includegraphics{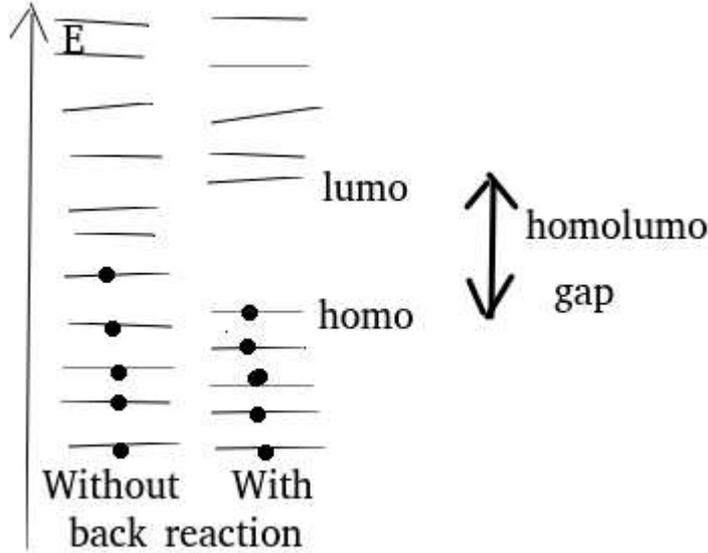}
\caption{\label{homolumo} Explanation of Homolumo-gap effect}
\end{figure}
        Let us consider the spectrum of energy levels for the electrons
        in a piece of material, e.g. one of our pearls, and at first assume
        that the positions or distributions of the charged particles in the material
        are fixed.

        Then the ground state is just a state built e.g. as a Slater
        determinant for the electrons being in the lowest single electron
        states, so many as are needed to have the right number of electrons.

        But now, if the charged particles can be moved due to their interactions, the
        ground state energy could be lowered by moving them so that the filled
        electron state levels get lowered.

        {\em So we expect introducing such a ``back reaction'' will lower
          the filled states.}


        When the filled levels get moved downwards,
         then the homo = ``{\bf h}ighest {\bf o}ccupied {\bf m}olecular
        {\bf o}rbit'' level will be lowered and its distance to the
        next level, the
        lumo (= {\bf l}owest {\bf u}noccupied {\bf m}olecular {\bf o}rbit),
        will appear extended on the energy axis, as illustrated in Figure \ref{homolumo}.

        {\em We believe that we can estimate the homolumo-gap $E_H$.}

        Using the Thomas-Fermi approximation - or crudely just some dimensional
        argument where the fine structure constant has the dimension of velocity -
        we calculated the homolumo gap in highly compressed ordinary matter for
        relativistic electrons \cite{theline}:
        \begin{eqnarray}
          E_H &\sim& (\frac{\alpha}{c})^{3/2}\sqrt{2}p_f\\
          \hbox{where } \quad p_f &=& \hbox{Fermi momentum} \\
          \frac{\alpha}{c}&=& \frac{1}{137.03...}
        \end{eqnarray}
        (the $\sqrt{2}$ comes from our Thomas-Fermi calculation).

        It is by requiring this homolumo-gap to be the 3.5 keV energy
        of the X-ray line mysteriously observed by satellites from clusters
        of galaxies, Andromeda and the Milky Way Center that we estimate
        the Fermi-energy to be $E_f \approx p_f =$ 3.3 MeV
        in the interior bulk of the pearl.


  {\bf Brief summary of theoretical ideas underlying our dark matter pearls}

  \begin{figure}
  \includegraphics[scale=0.9]{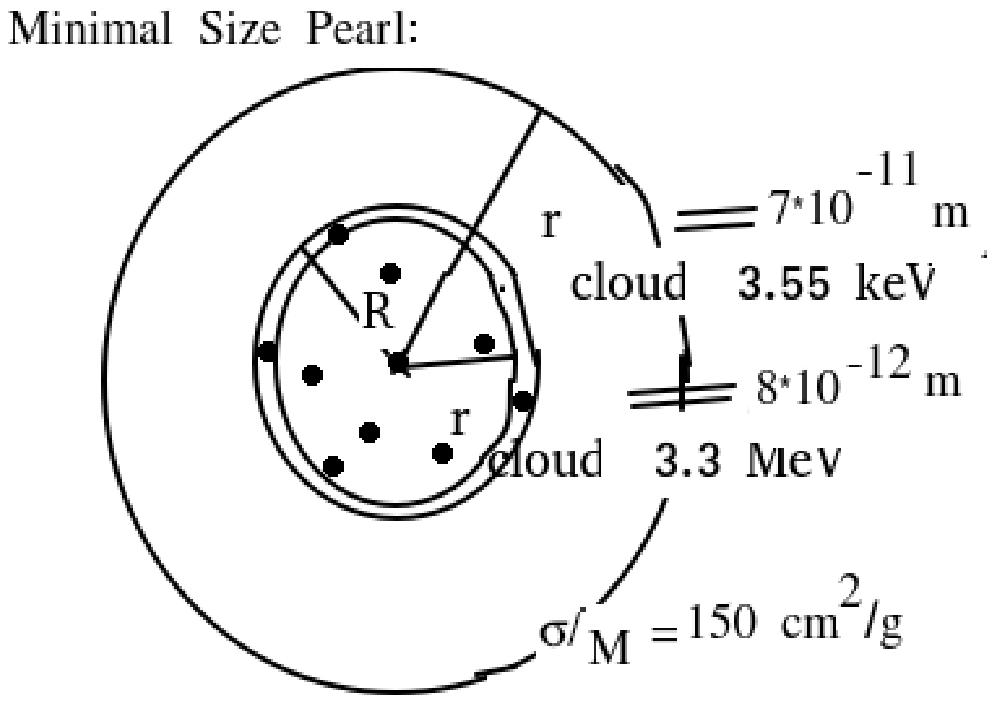}
\caption{ \label{Pearl2}Our Picture of Minimal Size Dark Matter Pearls.}
\end{figure}

\begin{itemize}
\item{\color{blue} Principle } Nothing but Standard Model!
  { (Seriously it would mean not in a BSM-workshop.)}
\item{\color{blue} New Assumption} Several Phases of Vacuum with Same Energy
  Density; this is the so-called Multiple Point Principle \cite{Corfu2017, Corfu2019, MPP1, MPP2, MPP3, MPP4, tophiggs, Corfu1995}.
\item{\color{blue} Central Part} Bubble of New Phase of Vacuum with e.g. carbon
  under very high pressure, surrounded by a surface with tension $S$
  (= domain wall) providing the pressure.
\item{\color{blue} Outer part} Cloud of Electrons much like an ordinary atom
  having a nucleus with a charge 
 $Z \approx 10^4$ to $10^{5}$.   
\end{itemize}

We provide our picture of  minimal size dark matter pearls in Figure
\ref{Pearl2}. 

\section{Non-gravitational Interactions}
\label{ngi}
The collisionless cold dark matter model provides a good description of the
large scale structure of the Universe. However there are various problems
at small scales \cite{annika, SIDM} for the hypothesis that dark matter only has
gravitational interactions. Originally Spergel et al \cite{firstSIDM}
suggested that the lack of a peak or cusp in the center of galaxy clusters, as
expected for cold dark matter with purely gravitational interactions, required
self interacting dark matter with a relatively large cross section. The
relevant parameter is in fact the cross section to mass ratio $\frac{\sigma}{M}$
and for the cores in galaxy clusters, where the collision velocity is
$v \sim$ 1000 km/s, a value $\frac{\sigma}{M} \sim$ 0.1 $cm^2/g$ is needed. The
self interaction can of course be velocity dependent and the cores in spiral
galaxies where $v \sim$ 100 km/s require  $\frac{\sigma}{M} \sim$ 1 $cm^2/g$. In
dwarf galaxies around our Milky Way, where dark matter moves more slowly
$v \sim$ 30 km/s, larger cross section to mass ratios
$\frac{\sigma}{M} \sim$ 50 $cm^2/g$ are needed.

  \begin{figure}
  \includegraphics[scale=0.9]{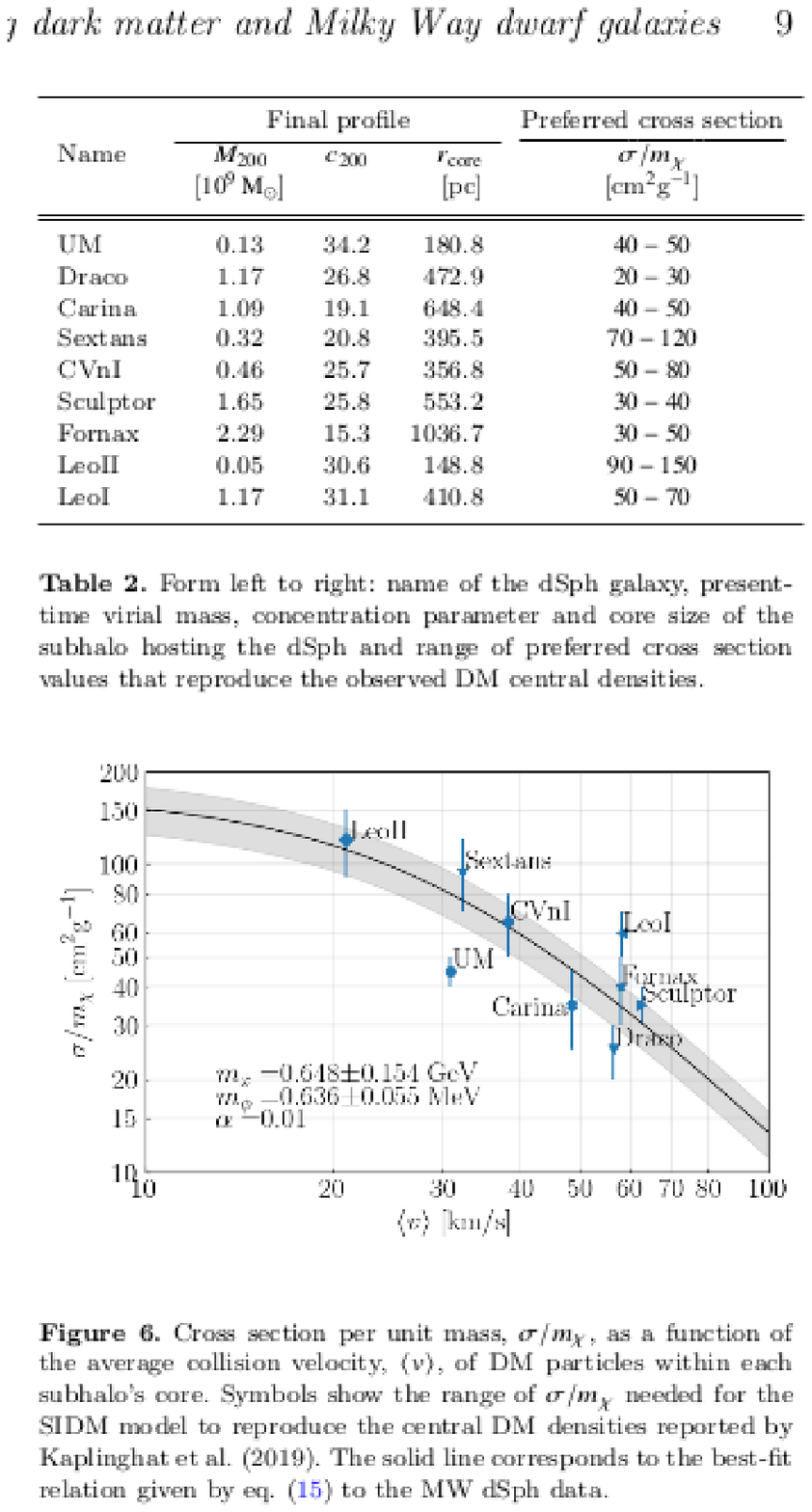}
  \caption{\label{Correa} Cross section to mass ratio $\frac{\sigma}{M}$ of self
    interacting dark matter particles as a function of the collision velocity
    $v$ in dwarf galaxies from reference \cite{CAC}.}
\end{figure}

  Recently Correa \cite{CAC} made a study of the velocity dependence of self
  interacting dark matter. In particular she analysed the Milky Way dwarf
  galaxies and her results are displayed in Figure \ref{Correa}. The
  extrapolation of Correa's fit to the data towards zero velocity points to
  the ratio $\frac{\sigma}{M} \rightarrow$ 150 $cm^2/g$. This ratio can be
  taken as an experimental estimate of the impact area over the mass as seen
  for very soft collisions. In our model the cross section in this low
  velocity limit is given by the extent out to which the electrons
  surrounding our pearls reach.

  
  \section{Minimal Size for the Pearls}
  \label{new}
It is our hope here to estimate the size of the pearls in our model under
the extra assumption that {\em the pearl size is the smallest allowed} using the
Fermi momentum $p_f = 3.3$ MeV of the electrons inside the skin of the pearl. 
The value of $p_f$ was determined in earlier papers \cite{Corfu2019, theline, Bled20} from the frequency and intensity of the 3.5 keV line emitted by various clusters of galaxies etc. Basically we want to show that there is
a certain thickness crudely of the electronic cloud  providing  
the electrical pressure needed to keep the surface tension skin spanned out.
If this needed thickness of the cloud, which really has a lack of electrons or
anti-cloud on the inner side of the pearl skin, should have more electrons than
the whole pearl, of course something would be wrong. Thus the radius of the
pearl must be at least so large as to allow enough protons in the pearl as to
make it possible to let a smaller number of electrons be pressed outside
and form an electric  attraction sufficient to carry the skin with its
tension spanned out.

\subsection{The Condenser of the Cloud and its Pressure on the Skin}

Have in mind that some electrons are pushed out of the skin in the philosophy
that the skin acts most strongly on the nuclei and not quite as forcefully
on the electrons; the electrons are highly degenerate, while the nuclei
need much less energy to achieve the possibility for the required quantum
fluctuation in the momentum. So we should imagine that near the skin there is
on the outside an excess of electrons while on the inside the corresponding
deficit of electrons. The full number of electrons must of course be equal to
that of the protons in the nuclei, since the total pearl must be essentially
neutral. Looking locally along the skin the electron cloud configuration
acts as a condenser with the negative electrons on the outside and the
missing electron positive charge on the inside. In such a condenser there
is of course an attractive force between the two plates (formed from the
electron and hole clouds). The expression for the force $F_{att}$ attracting the
plates in a condenser is
\begin{eqnarray}
  F_{att}&=& \frac{\epsilon_0 A V^2}{2d^2}\\
  &=& \frac{Q^2}{2A\epsilon_0}
\end{eqnarray}
where $A=4\pi R^2$ is the area of one of the condenser plates, which in our
case for thin clouds is the surface area of the pearl. The electric
tension of the condenser is here denoted $V$ and the distance between the
plates $d$ is imagined as small. The vacuum permeability is denoted
$\epsilon_0$.
Further $Q$ is the total charge on the condenser and $q=\frac{Q}{A}$
is the charge density per unit area. Writing the relation per unit area we have
\begin{eqnarray}
  \frac{F_{att}}{A} &=& \frac{q^2}{2\epsilon_0}\\
  &=& \frac{(\frac{Z_{outside}}{A})^2 *e^2}{2\epsilon_0}.\label{presparea}
\end{eqnarray}
where $Z_{outside}$ is the number of electrons outside the skin.

In our early work \cite{Tunguska} we required that the pressure
provided by the skin with tension $S$ and radius $R$ should be 
balanced by the relativistic electron degeneracy pressure
\begin{eqnarray}
  P&=& \frac{2S}{R}\\
  &\approx& \frac{1}{12\pi^2}*p_f^4.
\end{eqnarray}
So when electrons are pushed out of a layer inside the skin, the electron degeneracy pressure is replaced by the electric force per unit area
\begin{eqnarray}
 \frac{F_{att}}{A} &\approx & P = \frac{1}{12\pi^2}*p_f^4. \label{Fatt}
\end{eqnarray}

Now we have fitted to the intensity and the very line frequency of the
3.5 keV X-ray line emitted by galactic clusters \cite{Bled20} 
using only one parameter:
\begin{eqnarray}
  \frac{\xi_{fS}^{1/4}}{\Delta V} &=& 0.6\, MeV^{-1}.
\end{eqnarray}
The Fermi momentum of the electrons is given by this parameter to be 
\begin{equation}
   p_f= \xi^{-1/4}_{fS}2\Delta V = 3.3 \, MeV.
\end{equation}
  Thus the pressure is
\begin{eqnarray}  
   P &=&  \frac{1}{12\pi^2}*p_f^4\\
  &=& \frac{(3.3 \, MeV)^4}{12 \pi^2}\\
  &=& 1.00 \, MeV^4 \label{Press}
  \end{eqnarray}

Using that the fine structure constant is
\begin{eqnarray}
  \alpha&=& \frac{e^2/(\hbar c)}{4\pi \epsilon_0}
  \end{eqnarray}
 it follows from equations (\ref{presparea}), (\ref{Fatt}) and (\ref{Press}) that:
 \begin{eqnarray}
    2\pi \alpha (\frac{Z_{outside}}{A})^2 &=& 1.00 \, MeV^4
 \end{eqnarray}   
  giving
  \begin{eqnarray} 
    	\frac{Z_{outside}}{A} &=& \frac{1.00 \, MeV^2}{\sqrt{2\pi\alpha}}\\
&=& 1.17 *10^{26}m^{-2}.
\end{eqnarray}

Now the electron density in the undisturbed interior of our pearls is
\begin{eqnarray}
  n_e &=& \frac{1}{3\pi^2}*p_f^3\\
  &=& \frac{1}{3\pi^2}*(3.3 \, MeV)^3\\
  &=& 1.5*10^{38}m^{-3}.
\end{eqnarray}
So the layer thickness of material of electron number density
$n_e = 1.5*10^{38}m^{-3}$ having an area density of 
$\frac{Z_{outside}}{A} = 1.17*10^{26}m^{-2}$ is
\begin{eqnarray}\label{lt}
  \hbox{``layer thickness''}&=& \frac{1.17*10^{26}m^{-2}}{1.5*10^{38}m^{-3}}\\
  &=& 7.8 *10^{-13} m.
\end{eqnarray}

This then means that on the inside of the skin at least there must be
a range of the size of $\hbox{``layer thickness''} = 7.8 *10^{-13}m$
in which the electrons are removed relative to the usual density in the
deep inside of the pearl. But if the skin radius $R$ is not at least
of this size, it is nonsense. So we conclude that
\begin{eqnarray}
  R &\ge&\hbox{`` layer thickness''} = 7.8 *10^{-13}m
\end{eqnarray}
is the minimum size for a pearl with an electron Fermi momentum of 3.3 MeV.
 

This minimal size which we here found is clearly an underestimate of the
needed minimal size of the radius for the skin, since the electrons will
not be removed exactly in a layer with sharp edges. Therefore the
realistic minimal radius is some factor of order unity larger than our
number here $7.8 *10^{-13}m$. In fact the realistic limit cannot be far
away from the radius $5*10^{-12}m$ found in our Bled-proceedings 
contribution \cite{Bled21}.

\subsection{Little more thinking of the lower limit}

It is clear that even the possibility that the total number of electrons
inside the skin should be pushed out is absurd. There
must of course be a density (per unit volume) of electrons inside the skin of
at least the same size as in the outside region. So the lowering of the density
due to the missing electrons on the inside cannot be more than 50 \% of
the total density before the emptying out. This consideration would make the
$\hbox{``layer thickness''}$ two times as large. Taking into account that
in the case of minimal size we have to deal with a strongly curved skin compared to this
$\hbox{``layer thickness''}$, we also see that the inner thickness layer may
have to be about doubled to compensate for the smaller sphere area as one goes
deeper in. So overall we are close to having to scale up
this $\hbox{``layer thickness''}$ by a factor 4. That would bring us really
very close to the radius $5*10^{-12}m$ calculated in the Bled-proceedings paper
\cite{Bled21} and in the next section, see eq. (\ref{r3.3av}).

So we might take the point of view that we are thereby effectively calculating 
the minimal size of the pearl. Indeed in \cite{Bled21} and section \ref{range}
below, we assume that the number of electrons $Z$ outside the central part of 
the pearl is of order $\frac{M}{4m_N}$ meaning there are equally many electrons 
inside and outside as in the limiting case discussed above. 

\section{Range of Electron Cloud}
\label{range}
We can consider the calculation in this section as an estimate of the
smallest size of the pearl which is compatible with our parameter
$\frac{\xi_{fS}^{1/4}}{\Delta V} =0.6 \, MeV$. We also find
that the value of the cross section to mass ratio
$\frac{\sigma}{M}=150 \, cm^2/g$
determined from dwarf galaxies \cite{CAC} at the lowest velocity is compatible with this minimal size.

The range of the extension of electrons around the pearl is to first approximation (using a
Thomas Fermi approximation philosophy)
  supposed to be given by the requirement that the electron binding energy
  is of the order of the homolumo gap value 3.5 keV. So we denote this
  radius by $r_{cloud \; 3.5 keV}$.
  But probably the Thomas Fermi approximation, which we shall use, is not
  applicable to the very thinnest outskirts of the electron cloud. At the very
  outermost part of the electron cloud the more usual wave function thinking might be
  more applicable, and consequently the electron density will not
  fall off sharply at $r_{cloud \; 3.5 keV}$ as the Thomas Fermi approximation
  together with the homolumo-gap effect would suggest. Thus we should 
  rather believe in a somewhat bigger extension of the cloud,
   which compared to the
  use of a sharp cut off at $r_{cloud \; 3.5 keV}$ would give a larger
  low velocity cross section $\sigma$ and thus
  (for the same mass) a bigger ratio $\frac{\sigma}{M}$.
  
  Similarly the radius of the bubble
  containing the nucleons inside our dark matter pearl corresponds to a
  radius  $r_{cloud \; 3.3 MeV}$ at which the potential for the electron is
  -3.3 MeV (= Fermi energy of the electrons (numerically)). The high velocity
  hard collisions of our pearls, supposed to result in the unification of 
  two pearls into a single pearl,
  correspond to interactions between the bubble
  skins with a cross section of order $\pi r_{cloud \; 3.3 MeV}^2$.
  
\subsection{Thomas Fermi Approximation, Electron Cloud}  
  We will now consider the electric potential for our pearl using the
  Thomas-Fermi approximation for a heavy atom \cite{Thomas, Fermi, Spruch}.
  In this approximation the Coulomb
  potential of the ``nuclear" charge Z - which in our model should
  be the number of charge quanta outside the skin of
  radius $R=r_{cloud \; 3.3MeV}$ - is multiplied by the Thomas-Fermi
  screening function $\chi(r/b)$ where
 \begin{equation}
 b = 0.88 \frac{a_0}{Z^{1/3}}
 \end{equation}
 and $a_0$ is the Bohr radius.
 We assume, that the skin of the bubble or ``nucleus" of the pearl mainly
 acts on the {\em nucleons}
 or rather nuclei. So the electrons spread out and
 some of the electrons - in fact $Z$ - are outside the central part of
 the pearl inside the skin.
 Therefore $Z$ is also the effective charge of the central part of 
 the pearl or bubble of the new phase.

 In the Thomas-Fermi approach we are then led to the following equations
 for $r_{cloud \; 3.5 keV}$ and $r_{cloud \; 3.3 MeV}$:
\begin{eqnarray}
  \frac{\alpha *Z}{r_{cloud  \; 3.5 keV}}*\chi(r_{cloud \; 3.5 keV}/b)&=& 3.5 \; keV
  \label{3.5keV}\\
   \frac{\alpha *Z}{r_{cloud  \; 3.3 MeV}}*\chi(r_{cloud \; 3.3 MeV}/b)&=& 3.3 \; MeV
  \label{355keV}\\
  b&=& 0.88*\frac{a_0}{Z^{1/3}}\label{bdef}\\
\end{eqnarray}
We identify $r_{cloud  \; 3.5 keV}$ with the radius of the electron
cloud and $r_{cloud \; 3.3 MeV}$ with the skin radius $R$ of the pearl.

If we decide to look for the minimal size pearl, an assumption that seems
to be approximately fulfilled\footnote{However we shall consider much bigger
pearls in section \ref{discuss} in order to be consistent with an annual
modulation in the DAMA-LIBRA data.}, 
we can add to the above set of equations
the assumption that {\em the number of electrons outside the skin and
  inside the skin are of similar order of magnitudes}. To be a bit more precise
we can say: Thinking of it as at outset that the electrons were all kept
inside, like the nuclei, we must have removed a number of electrons from the
inside equal to the amount present in the outside cloud, because otherwise
the full pearl would not be neutral. But it would be quite unreasonable that
there should be a lower electron density in the inside than in the
outside. So approximating for a moment the skin by being flat we can at
the very most have half of the electrons in the outside. Since we assume that
the nuclei inside the skin are so light as to have 
roughly equally many protons and neutrons we can, remembering that the
nuclei dominate the mass M, write the above result as an inequality:
\begin{eqnarray}\label{ieqMZ}
  M/m_N &\ge& 4Z,
  \end{eqnarray}
where $Z$ is the number of electrons outside the skin in the cloud.
The assumption of the minimal size becomes then that there is an
equality instead of the inequality (\ref{ieqMZ}),
\begin{eqnarray}\label{eqMZ}
  M/m_N &\approx & 4Z \hbox{ for minimal size.}
  \end{eqnarray}

It is going to be a success of our model that
using this minimal size assumption 
we get a similar
value to the Thomas-Fermi value (\ref{355keV}) for $R \approx r_{cloud \; 3.3 MeV}$ using another
method to calculate it. We shall use
\begin{equation}
\frac{\sigma}{M}|_{v \rightarrow 0} = 150 \; cm^2/g
\end{equation}
and
\begin{equation}
\sigma = \pi * r_{cloud \; 3.5 keV}^2
\end{equation}
to determine the mass M. Then using the formula for the mass of a pearl in
terms of the radius $R$ and the Fermi momentum \cite{theline, Bled20}
\begin{equation}
\frac{M}{m_N}=\frac{8}{9\pi} * (R*p_f)^3,\label{Mpf}
\end{equation}
we can calculate another value for $R$.

In our updated contribution to the Bled Proceedings from 2020 \cite{Bled20}
we estimated a pearl mass\footnote{However we now prefer a value of 
$M\sim 10^{12} \, m_N$; see section \ref{parameters}.} of $M \sim 10^5$ GeV.
So we take here $Z=5.3*10^4$ (mainly from the ``historical value'' under
the assumption of minimal size) 
as a typical charge in the central part of the
pearl, for which then $b = 1.24 * 10^{-12} m$. Using numerical values for the
Thomas-Fermi screening function in the
paper \cite{Parand}, we obtain from (\ref{3.5keV}) the radius of the
electron cloud to be
\begin{equation}
r_{cloud \; 3.5 keV}=
4.96*10^{-11}m. \label{3.5k}
\end{equation}Then assuming - from the dwarf galaxies observations - the low velocity
ratio $\frac{\sigma}{M}=150 \; cm^2/g$ we obtain
\begin{eqnarray}
  M&=& \frac{\pi *(
    4.96*10^{-11}m)^2}{150 \; cm^2/g}\\
  &=&5.2*10^{-19}g\\
  &=&3.1 *10^5  \; m_N \label{Bledmass}
  \end{eqnarray}

As a side remark notice that, using our proposed minimal-size-rule of taking
$Z$ to be a quarter of the number $M/m_N$, we would get
$Z=8*10^4$ to be compared with our input here $5.3*10^4$, which is very well
consistent within a factor 2.

Next using (\ref{Mpf}) with $p_f =3.3 \; MeV = 1.6 *10^{13} m^{-1}$

\begin{eqnarray}
  (R*p_f)^3 &=&
  3.1*10^5*\frac{9\pi}{8}\\
   &=&
  10.9*10^5
  \end{eqnarray}
  giving
 \begin{eqnarray}
   R &=& \frac{\sqrt[3]{10.9*10^5}}{1.6*10^{13}m^{-1}}\\
 &=& 6.4*10^{-12}m.\label{r3.3M}
  \end{eqnarray}
This is to be compared with the Thomas-Fermi value obtained from
(\ref{355keV}) using the numerical values for $\chi(r/b)$ in \cite{Parand}
  \begin{eqnarray}
   R = r_{cloud \; 3.3 MeV} &=&
  3.66*10^{-12}m. \label{r3.3TF}
  \end{eqnarray}
  These two different estimates of the radius $r_{cloud \; 3.3MeV}$ at which the
  potential is 3.3 MeV essentially coincide to the accuracy of our
  calculation; they deviate by a factor of order unity 6.4/3.7 =1.7. So we
  could claim that formally our model is able to predict the low velocity
  limit $\frac{\sigma}{M}|_{v \rightarrow 0}$ in agreement with the value $150 \; cm^2/g$
  estimated from the study of dwarf galaxies around the Milky Way.

  We shall take the average of the two values  (\ref{r3.3M}) and (\ref{r3.3TF}) as
  our best estimate of the bubble skin radius, under the assumption of
  ``minimal size'':
  \begin{equation}
  r_{cloud \; 3.3MeV} = 5.0 *10^{-12} m \hbox{ (assuming ``minimal size'')}
  \label{r3.3av}
  \end{equation}
  and from (\ref{3.5k}) we have the radius of the electron cloud
  \begin{equation}
  r_{cloud \; 3.5keV} = 5.0 *10^{-11} m.\hbox{ (assuming ``minimal size'')}
  \end{equation}
  We note that these two radii differ by an order of magnitude, which means
  that the quantity $\frac{\sigma}{M}$ for our
  minimal size pearls should differ by {\em two
  orders of magnitude} between low velocities and high velocities, as
  astronomical observations indicate is the case for self interacting dark
  matter \cite{CAC}.


\section{Correction to have
	$``\chi"(x_{cloud \; 3.3 MeV})(1-\frac{xd\chi}{\chi dx})=1$ at $r=R$}
\label{correction}
In the previous section we used
the Thomas-Fermi screening function $\chi(r/b)$ satisfying 
the boundary condition that
$\chi(0)=1$ corresponding to a nucleus of zero radius compared to the
atomic size. This approximation is good for real atoms, and it is the
solution of the Thomas-Fermi differential equation with this 
boundary condition that has been studied and calculated.
However, our genuine pearl inside the skin is not so terribly small compared to
the atom or in our case the cloud of electrons. We therefore should in principle
do a better job by imposing the boundary condition
$\chi=1$ at that value of the variable $x =r/b$,
which corresponds to the radius of the skin $R=r_{cloud \; 3.3 MeV}$.

Now, however, what boundary condition to take is a little bit more
complicated: At the boundary value of $r=R$ at which we start the cloud
calculation we have the charge $eZ$ inside the sphere of this
radius $R$. That means that the {\em electric field} radiating from
the sphere is given by $eZ$, but the {\em potential} (as the reader
should have
in mind is gauge dependent and only defined by say the potential at
infinity being fixed to 0) does not only depend on the charge inside
the sphere, but also on what is outside, if the gauge is fixed in the
outside, namely at infinity.

The $\chi$-function with boundary condition at $r=R$ rather than at $r=0$
is a different function from the ``universal function'' $\chi$ as
evaluated in say the Parand article \cite{Parand}, but let us here
denote it by the letter $\chi(x)$ anyway, and just remember when we use
which boundary condition.

Considering for a moment $\chi$ as a function of the distance to the center
$r$ we obtain for the potential
\begin{eqnarray}
	V_{eff}(r) &=& \frac{e^2Z \chi}{r}
\end{eqnarray}	
	the radial field  
\begin{eqnarray}	
	eE&=&\frac{dV_{eff}(r)}{dr}\\
	&=& e^2Z\frac{-\chi + r\frac{d\chi}{dr}}{r^2},
\end{eqnarray}	
which should equal the Coulomb field 
\begin{eqnarray}
 eE_{Coulomb}&=&
	-e^2Z\frac{1}{r^2}
\end{eqnarray}	
Thus at $r=R$ we have the boundary condition:  
\begin{eqnarray}
-1&=& -\chi+r\frac{d\chi}{dr}
\end{eqnarray}
meaning 
\begin{eqnarray} 
1&=&\chi *(1-\frac{rd\chi}{\chi dr})\\
	&=& \chi *(1-\frac{xd\chi}{\chi dx}). \label{condition}
\end{eqnarray}
 
We found that the
$x$-variable value corresponding to the radius $r_{cloud \; 3.3. MeV}$, where the
potential equals the Fermi energy in the bulk of the interior skin
material in our model, namely 3.3 MeV, is $x_{cloud \; 3.3 MeV}= 2.95$
(actually using the boundary at $x=0$, so it is only crude). 
Near this value $x=2.95$ the $\chi$-function (with boundary condition at
$x=0$ as usual) is approximated by being proportional to
the power-law
\begin{eqnarray}
	\chi(x) &\propto& x^{-1.2} \quad \hbox{ for $x$ near 2.95 }
\end{eqnarray}	
and the value is 
\begin{eqnarray}
\chi(2.95) &=& 0.16.
\end{eqnarray}
Thus 
\begin{eqnarray}
 \chi(2.95)(1-\frac{x d\chi(x)}{\chi(x)dr})
	&=& 0.16*(1 + 1.2)\\
	&=&0.35
\end{eqnarray}

To get a crude estimate we take it that we can find a solution of the Thomas-Fermi 
differential equation, which
we avoid calculating explicitly, satisfying the boundary condition (\ref{condition}) 
by scaling the $\chi$ up by the inverse of the value
0.35 at the point $r = R$.  
Since it is
a non-linear differential equation we should solve it again, it is
in truth more complicated, but scaling by
$0.35^{-1}=3$ would then roughly
mean that a given value is taken on for a larger $x$ by a factor $f$ such
that $f^{(p+1)} = 3$. 
The power $p$ is a power chosen so that in the region of interest here
between the two radii $r_{cloud \; 3.3 MeV}$ and $r_{cloud \; 3.5 keV}$, the
function $\chi$ can be considered proportional to $x^{-p}$. Let us crudely
estimate this power $p$, for simplicity taking the corresponding 
range of $x$-values as $3$ and $30$.
The power $p+1$ used  in the formula $f^{(p+1)}=3$ was due to the 
{\em potential }
going  proportional to $\chi/r \propto \chi/x$. Now we have
\begin{equation}
	\chi(3) =0.1566 \quad \hbox{and} \quad \chi(30) =2.255*10^{-3}.
\end{equation}
This gives a fall over one decade of 
\begin{equation}
	``fall" = \frac{0.1566}{0.002255} = 69.4.
\end{equation}
So we obtain
\begin{equation}
	p = \log(``fall") =1.84.
\end{equation}
That is to say the factor is 
$f=
\sqrt[2.84]{3}=1.47 $. So very crudely the
radius say at which the potential is 3.5 keV and where the electron cloud
should end, would be pushed out to a bigger radius by a factor
1.47.
\begin{eqnarray}\label{r35corr}
	r_{cloud \; 3.5 keV}=5*10^{-11}m  &\rightarrow& r_{cloud \; 3.5 keV}=
	1.47*5*10^{-11}m\\ 
	&=&
	7.4*10^{-11}m. 
\end{eqnarray}

\subsection{Also the Skin Radius gets Corrected}

For the analogous correction of the radius of the skin $R=r_{cloud \; 3.3 MeV}$
we should use the power $1.2$ meaning approximating the
$\chi$ to be proportional to $x^{-1.2}$, because we work very close
to where this power is o.k.. Then the increasing of the value of the
$\chi$ function by a factor $3$ means that the value at which the
potential has a required value $3.3 MeV$, will go up by a factor
$f$, where now this $f$ must obey $f^{1+1.2}=3$. This means a correction factor
for the radius $r_{cloud \; 3.3 MeV}$ of $f =\sqrt[2.2]{3} =1.65$ giving
\begin{eqnarray}
	r_{cloud \; 3.3MeV} = 5*10^{-12}m &\rightarrow & r_{cloud \; 3.3 MeV}=
	1.65*5*10^{-12}m\\
	&=& 8.2*10^{-12}m\label{corrRmin}
\end{eqnarray}

It follows from equation (\ref{Mpf}) that the mass density of the pearl is
given by
\begin{eqnarray}
	\rho_B &=& \frac{2m_N}{3\pi^2}p_f^3 \\
	&=& \frac{2}{3\pi^2}(940 \, MeV)(3.3 \, MeV)^3 \\
	&=& 5.2*10^{11} \, kg/m^3. \label{rhoB}
\end{eqnarray}

Using the density (\ref{rhoB}) of the part of the pearl inside the skin
we obtain the mass of the pearl to be
\begin{eqnarray}
  M	&=& \frac{4\pi}{3}*5.2*10^{11}kg/m^3 * (8.2*10^{-12}m)^3\\
	&=&  1.19*10^{-21} \, kg\\
	&=& 7.0*10^{5} \, GeV\label{corrMmin}
\end{eqnarray}

Here we just improved the calculation for the radius $R$ of the minimal size pearl
and thus got the minimal mass to be $M=7.0*10^{5}GeV$.
So we obtain
\begin{eqnarray}
   \frac{\sigma}{M}=   \frac{\pi r_{cloud \; 3.5 keV}^2}{M} &=& 
   \frac{\pi (7.4*10^{-11})^2}{7.0 GeV*1.78*10^{-27} kg/GeV}\\
   &=& 140 cm^2/g,	
\end{eqnarray}
still in agreement with the low velocity estimate $\frac{\sigma}{M} \simeq 150cm^2/g$ from the study \cite{CAC} of the Milky Way dwarf galaxies. 



\section{Achievements}
\label{Achievements}


\begin{itemize}
\item{{\bf \color{blue} Low velocity $\frac{\sigma}{M}|_{v\rightarrow 0}$ cross
    section to mass ratio.}} The a priori story, that dark matter has only
  gravitational interactions seems not to work perfectly: Especially in dwarf
  galaxies (around our Milky Way) where dark matter moves relatively slowly
  an appreciable self interaction
  cross section to mass ratio $\frac{\sigma}{M}$ is needed.
  According to the fits in \cite{CAC} this ratio has the low velocity limit
  $\frac{\sigma}{M}|_{v\rightarrow 0} = 150 \;cm^2/g$.
  We may say our pearl-model
  ``predicts'' this ratio in order of magnitude provided that the
  pearls are of ``minimal
  size'', meaning that the pearls are as small as possible while
  it is still true that there can be a sensible diminishing of electron
  number in the inside compared to that we had with completely neutral material 
  inside.
  










\item{{\bf \color{blue} Can make the Dark Matter Underground Searches get
    Electron Recoil Events}} Most underground experiments are designed to look
  for dark  matter particles hitting the nuclei in the experimental apparatus,
  which is then scintillating so that such hits presumed to be on nuclei can
  be seen.
  But our pearls are excited in such a way that they send out energetic
  {\em electrons} (rather than nuclei) and this does not match with what is
  looked for, except in the DAMA-LIBRA experiment. In this experiment
  the only signal for
  events coming from dark matter is a seasonal variation due to the Earth
  running towards or away from the dark matter flow.



  \item{\bf \color{blue} The Intensity of 3.5 keV X-rays from Clusters etc.}
    We fit the very photon-energy 3.5 keV and the overall intensity from a
    series of clusters, a galaxy, and the Milky Way  Center \cite{Bled20} with
    one parameter
    $\frac{\xi_{fS}^{1/4}}{\Delta V} = 0.6 \; MeV^{-1}$.

  \item{\bf \color{blue} 3.5 keV Radiation from the Tycho Supernova
    Remnant.} Jeltema and Profumo \cite{Jeltema} discovered the 3.5 KeV
    X-ray radiation coming
    from the remnant of Tycho Brahe's supernova, which was unexpected for
    such a small source. We have a
    scenario giving the correct order of magnitude for the observed intensity
    in our pearl model: supposedly our pearls are getting excited by the
    high intensity of cosmic rays in the supernova remnant \cite{Bled20}.


   Even though we only need the one parameter
    $\frac{\xi_{fS}^{1/4}}{\Delta V}=\frac{2}{p_f}$, it is nice to know the
    notation:

  \begin{eqnarray}
    \Delta V &=& \hbox{`` difference in potential for a nucleon between the}
    \nonumber\\
    && \hbox{inside
      and the outside of the central part of the pearl''}\nonumber\\
    &\approx & 2.5 \; MeV\\
    \xi_{fS}&=& \frac{R}{R_{crit}} \quad \hbox{estimated to be } \approx 5\\
    \hbox{where } R &=& \hbox{``actual radius of the new vacuum part''}
    \nonumber\\
    &\approx & r_{cloud \; 3.3 MeV}\\
    \hbox{ and } R_{crit}&=& \hbox{`` Radius when pressure is so high}\nonumber\\
    && \hbox{ that nucleons
      are just about being spit out''}
    \end{eqnarray}
  The subscript $fS$ on the parameter $\xi_{fS}$ indicates that the surface
  tension $S$
is fixed independent of the radius $R$.


  \item{\color{blue}DAMA-rate} Estimating observation rate of DAMA-LIBRA
    from kinetic
    energy of the incoming dark matter as known from astronomy.
  \item{\color{blue}Xenon1T Electron recoil rate} Same for the
    {\em electron recoil
    excess} observed by the Xenon1T experiment.
  \end{itemize}
In order to explain these last calculational estimates it is necessary to
know how we imagine
the dark matter to interact and get slowed down in the air and the earth
shielding; also how the dark matter particles get excited and emit
3.5 keV radiation
  or electrons.

  {\bf About the Xenon1T and DAMA-rates:}
  \begin{itemize}
  \item{\color{blue} Absolute rates very crudely} Our estimate of the
    absolute rates for the
    two experiments are very very crude, because we assume that the dark matter
    particles - in our model small macroscopic systems with tens of thousands
    (for the larger pearls considered in section \ref{dirtypearl} rather
    $\sim 10^{12}$)
    of nuclei inside them - can have an exceedingly smooth distribution of
    lifetimes on a logarithmic scale.
     These calculations are discussed in section \ref{sec5}.

  \item{\color{blue} The ratio of rates} The ratio of the rates in the
    two experiments  - Xenon1T electron recoil excess  and DAMA - should in
    principle be very accurately predicted in our model, because they are
    supposed
    to see exactly the same effect just in two different detectors in the
    same underground laboratory below the Gran Sasso mountain!
    One would therefore expect the rates to be the same, but the Xenon1T rate
    is 250 times smaller than the DAMA rate. We briefly refer to a possible
    resolution of this problem, which needs further study, in
    section \ref{sec5}.
    \end{itemize}
\section{Impact}
\label{Impact}
  {\bf Illustration of Interacting and Excitable Dark Matter Pearls}

  \includegraphics[scale=0.6]{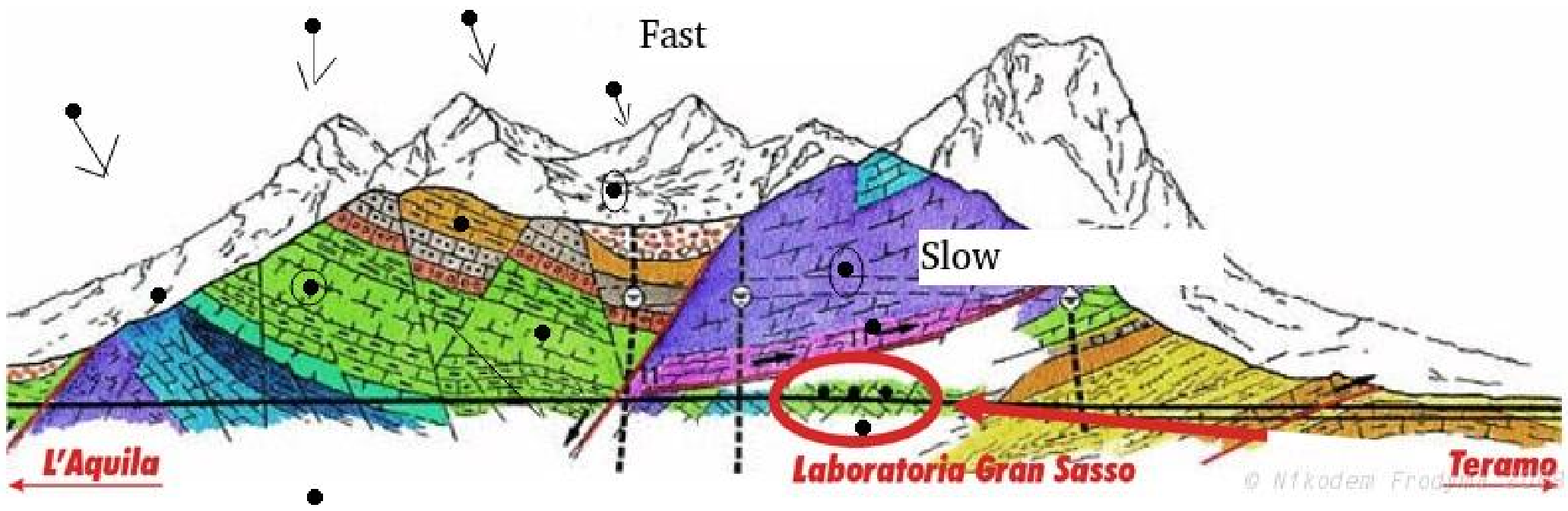}

  The dark matter pearls come in with high speed (galactic velocity), but
  get slowed down to a much lower speed by interaction with the air and
  the shielding
  mountains, whereby they also get excited to emit 3.5 keV X-rays or
  {\em electrons}. We shall consider two types of pearl here: the minimal 
  size pearl with radius $R \sim 8*10^{-12} \, m$ discussed so far and 
  a pearl introduced in section \ref{dirtypearl} of 
  radius $R \sim 10^{-9} \, m$, which is 
  surrounded by dirt in outer space that is quickly burnt off in the 
  atmosphere. 

  \medskip

  {\bf Pearls Stopping and getting Excited in Earth Shield}
  
  What happens when the dark matter pearls in our model
  hit the earth shielding above the experimental halls of e.g. DAMA?

  \begin{itemize}
  \item{\color{blue} Stopping} Taking it that the pearls stop in
    the earth: The pearls are slowed down in about $5*10^{-6}s$
    from their galactic speed of about 300 km/s down to a 
    speed $\sim 49$ km/s below which
    collisions with nuclei can no longer excite the $3.5 \; keV$ excitations.
    Using the value of $\frac{\sigma}{M} \sim 2 \, cm^2/g$ obtained in the 
    fit to the dwarf galaxies in \cite{CAC} for the minimal size pearl we 
    get the stopping length to be $d \sim 2/3 \, cm$. On the other hand for the pearl of radius 1 nm we get a stopping length $d \sim 22 \, cm$. 

    But taking it that they stop in the air, which is more likely:
    They are slowed down over a range of about 7 km - as the density of the
    atmosphere goes up by a factor $e=2.71..$ over such a range - in about
    $2*10^{-2}s$. The above stopping lengths in earth correspond in 
    the penetration into the atmosphere to stopping at the heights
    \begin{eqnarray}
    	h = 44 \, km &\hbox{for}& \hbox{minimal size pearls}\\
    	h = 19 \, km &\hbox{for}& \hbox{pearls with} \, R \sim 1 \, nm.
    \end{eqnarray} 
     
  \end{itemize}

  \begin{itemize}
  \item{\color{blue}Excitation}
    As long as the velocity is still over ca. 49 km/s, collisions with
    nuclei in the
    shielding can excite the electrons inside the pearl by 3.5 kev or more
    and make pairs of quasi electrons and holes say. We expect that often the
    creation of (as well as the decay of) such excitations require electrons to
    pass through a (quantum) tunnel and that consequently there will be
    decay half lives of
    very different sizes. We hope even up to many hours or days or years ...



  \item{\color{blue} Slowly sinking:}
    After being stopped in of the order of 2/3 cm of the shielding
    for a minimal size pearl or 22 cm for a 1 nm size pearl,
    the pearls continue with a much lower velocity driven by the gravitational
    attraction of the Earth.
    If a high Reynolds number formula would have been valid the pearls might
    manage to come down to the experimental halls in days. However
    if - as we now believe -
    they are so slow as to rather make Stokes law be used it becomes a
    problem to get them down inside a year, as we require since
    otherwise the DAMA-seasonal-effect tends to be smeared away.

    When the pearls after hopefully less than a year
    reach down the
    1400 m to the laboratories, most of the pearls have returned to their
    ground states, but some exceptionally long living excitations survive.
  \end{itemize}

   {\em Note that the slowly sinking velocity is so low that
    collisions with nuclei cannot give such nuclei enough speed to excite the
    scintillation counters neither in DAMA nor in Xenon-experiments.}


  \begin{itemize}

  \item{\color{blue} Electron or $\gamma$ emission}
    Typically the decay of an excitation could be that a hole in the Fermi sea
    of the electron cloud of the pearl gets filled by an outside electron
    under emission of another electron by an Auger-effect. The electron must
    tunnel into the pearl center. This can make the decay lifetime
    become very long
    and very different from case to case.
    \end{itemize}


  {\bf Emission as electrons or photons makes Xenon-experiments not see events,
    except...}

  That the decay energy is released most often as electron energy meaning
  that such events
  are discarded by most of the Xenon-experiments, which only expect the nucleus
  recoils to be dark matter events. This would explain the long standing
  controversy
  consisting in DAMA seeing dark matter with a much bigger rate than the upper
  limits from the other experiments.

  Rather recently though Xenon1T looked for potential excess events among the
  {\em electron recoil events} and found $16 \; events/year/tonne/keV$
  in the lowest keV-bands over a background of the order of
  $(76 \pm 2) \; events/year/tonne/keV$.

  In our model this rate should be compatible with the DAMA event rate.
  However they deviate by a factor of 250. It therefore appears that we need
  the pearls to run much faster through the xenon-apparatus than through the
  DAMA one.

  \section{Consistency of the Picture of Slowed Down Impact Pearls}
\label{discuss}
  We postulate - as strongly indicated from the range of energies seen
  in the two experiments being both concentrating around 3.5 kev -
  that the Xenon1T electron recoil excess and the DAMA effects should be due
  to the same effect, namely that dark matter particles although stopped
  or slowed down de-excite in the detectors and emit energy in quanta
  of 3.5 keV. Then it seems, as we just said, that the rate per kg 
  of scintillator say
  should be rather closely the same in the two experiments. But now the
  DAMA-experiment sees 250 times as many events - even counting only the
  modulated part of the signal - as Xenon1T sees as excess electron 
  recoil events.

  One could of course speculate that the DAMA experiment sees some sort of
  emission not visible in the Xenon1T experiment, but we prefer the
  explanation that the pearls run faster through the liquid xenon than the
  solid NaI and thus spend less time in the xenon-detector than in the
  NaI one per kg.

  We shall now look at what is required to make such a picture work and be
  consistent:

  \subsection{Need for reaching down in less than one year}

  Since the DAMA experiment sees the seasonal variation of the signal,
  it is of course needed that the pearls reach down to the experimental halls
  faster than or at least not much slower than in one year. If the pearls reach
  down much slower than in a year the signal would be smeared out in time
  and the modulated signal would almost be washed away.

  But now most of the way down through the shielding there is solid
  material, not exactly NaI, but something which is similar to that
  solid.

  If we shall explain the ratio of 250 times more particles or events in
  DAMA compared to the Xenon1T electron recoil excess by the slower passage
  in the NaI than in the liquid xenon, then we need that the terminal velocity
  - when the drag force just balances gravity - is 250 times smaller in NaI
  than in the liquid xenon. But to avoid the modulation being smeared out
  the strongly slowed down pearls should at least go through the shielding 
  to the detectors in less than one year. The shielding
  in Gran Sasso contains 1400 m of mountain, and in addition the pearls may be
  slowed down already in the atmosphere, exactly how high in the atmosphere is
  of course dependent on the parameters of the pearls, the cross section 
  and mass. However the time for the pearls to pass through the atmosphere 
  will be much smaller than the time to pass through 1400 m of stone
  and can be safely ignored.

  For the pearl to reach through 1400 m in one year requires its
  velocity to be at least
  \begin{eqnarray}
    v_{needed} &=& \frac{1400m}{1 \, year}\\
    &=& 4.43 *10^{-5}m/s.
  \end{eqnarray}
  In the liquid xenon the pearl should fall with a 250 times
  larger speed, i.e.
  \begin{eqnarray}\label{needLXe}
    v_{terminal \; liquid \; Xe}&\ge & 250*4.43*10^{-5}m/s=1.11*10^{-2}m/s.
    \end{eqnarray}

  \subsection{Size requirement from Speed Requirement}

  Using Stoke's law for a ball with mass 
  \begin{eqnarray}
    M &= & \rho_B *\frac{4}{3}\pi R^3
  \end{eqnarray}
  one finds its terminal velocity in a gravitational
  field of strength $g$ to be
  \begin{eqnarray}
    v_{terminal}&=& \frac{2g\rho_BR^2}{9\eta},
    \end{eqnarray}
  where $\eta$ is the dynamical viscosity of the fluid through which the ball
  falls. Here $\rho_B$ is the density of the ball and    $R$ its radius.
  In this formula the ball is taken to be homogeneous, so it means for our pearls
  that one has ignored the surrounding electron cloud, which has in the model an
  essentially zero density, and the radius $R$ is the radius of the
  skin preferably.

  Hence in our model we rather have an inequality
  \begin{eqnarray}
     v_{terminal}&\le& \frac{2g\rho_BR^2}{9\eta}
    \end{eqnarray}
  where the equality will be more and more true the bigger the pearl.

  Inserting the minimal required value of the terminal velocity in liquid xenon (\ref{needLXe})
  of $v_{terminal} = 1.11 *10^{-2}m/s$ in our inequality we obtain
  \begin{eqnarray}\label{ineq}
    1.11*10^{-2}m/s &\le&  \frac{2g\rho_BR^2}{9\eta},
    \end{eqnarray}
  where of course now the dynamic viscosity $\eta$ should be the one for
  liquid xenon. For our crude calculations we use instead
  the data on the viscosity of gaseous and liquid nitrogen 
  \cite{nitrogen} to estimate a typical 
  value for the dynamical viscosity of the fluid form of a normal gas near its
  boiling point to be
  \begin{eqnarray}
    \eta_{liquid \; gas} 
    &\approx& 100 \mu Pa \, s = 0.1mPa \, s \quad \hbox{ near the boiling point. }
    \end{eqnarray}

  We take the density of the material inside the skin from eq. (\ref{rhoB}) 
  \begin{eqnarray}
    \rho_B&=& 5.2*10^{11}kg/m^3.
  \end{eqnarray}
  The inequality (\ref{ineq}) then means
  \begin{eqnarray}
    R^2&\ge&\frac{1.11*10^{-2}m/s*9\eta}{2g\rho_B}\\
    &=& \frac{1.11*10^{-2}m/s*9 *10^{-4}Pa s}{2*9.8m/s^2*5.2*10^{11}kg/m^3}\\
    &=& 0.98*10^{-18}m^2
    \end{eqnarray}
   giving 
   \begin{eqnarray}
    R&\ge&
    0.99*10^{-9}m. \label{minRyear}
  \end{eqnarray}
  The mass corresponding to this borderline radius
  $R=0.99*10^{-9}m$ is
  \begin{eqnarray}
    M_{minimal} &=& \frac{4\pi}{3} \rho_B*R_{minimal}^3\\
    &=& \frac{4\pi}{3}*5.2*10^{11}kg/m^3*
    (0.99*10^{-9}m)^3\\
    &=&
    2.1 *10^{-15}kg\\
    &=&
    1.2 *10^{12} \, GeV.\label{minMyear}
    \end{eqnarray}
  This is to be compared with our fit involving the  
  ``minimal size" pearls (\ref{corrMmin}) with
  $\sigma/M \sim 150 \, cm^2/g$, 
  which gave $M=
  7.0*10^5 \, m_N$.

  We have formally here an inconsistency of the order
  $\frac{1.2*10^{12}m_N}{7.0*10^5 m_N} =1.7*10^6$. So, in order  
  to avoid wash out of the seasonal effect in DAMA and to have enough difference
  in the speed of the pearls in the fluid xenon and the shielding, we need
  $
  1.7*10^6$ times as big mass $M$ of the pearl as we get from fitting the
  dwarf galaxy number $\sigma/M \sim 150 \, cm^2/g$.

  \subsection{Hope for better fit by using $v\sim 300 \, km/s$ instead of $0$}
  \label{hope}
  The discrepancy between the above two mass estimates might be resolved 
  if instead of the
  zero velocity limit $\frac{\sigma}{M} =150 \, cm^2/g$ we could use a 
  smaller value.
  We namely in our attempt to fit
  the ratio assumed that the electron cloud of the pearl only extended out to the
  distance $r_{cloud \; 3.5 keV}$ where the electron potential is 3.5 keV, as
  expected from the homolumo gap. But
  it is known that one cannot trust the Thomas Fermi approximation completely,
  even if the homolumo gap is working, so as to cut off the density of
  electrons at this $r_{cloud \; 3.5 keV}$ distance {\em sharply}. Rather there will be a tail of the electron cloud even further out, 
  and at a true zero velocity even
  such a very thin cloud of electrons could cause some cross section.
  However, it is rather sure, that if we consider pearls meeting each other
  with a relative velocity $v$ such that the kinetic energy of the pearl
  $\frac{1}{2}Mv^2$ taken per electron in the cloud, i.e.
  $\frac{Mv^2}{2Z} \approx \frac{Mv^2}{2 *\frac{1}{4}M/m_N}$
  = $2m_Nv^2$ (if we take $Z\approx M/(4m_N)$ as in the minimal size case),
  becomes equal to the binding or gap energy 3.5 keV, then at that velocity the
  one pearl will penetrate into the other to the potential depth of this order.
  So at the velocity given by
  \begin{eqnarray}
    \frac{1}{2}*4m_N v^2 &\sim& 3.5 \, keV\\
    \hbox{giving } \quad v&\sim& \sqrt{\frac{0.0035 MeV}{2*940MeV}}\\
    &=& 0.00136c\\
    &=& 4.1 *10^5 \, m/s = 410 \, km/s
  \end{eqnarray}
  the kinetic energy per cloud electron is 3.5 keV.
  Extrapolating the curve in Figure 5 for $\frac{\sigma}{M}$ as a function
  of the velocity $v$ to 410 km/s gives about $1 \, cm^2/g$ for the ratio.
  The speculation would then be that for lower velocities
  the ``tail'' distribution, which should not have been there in 
  the Thomas Fermi approximation with the homolumo gap cut off,
  is important and causes a bigger cross section. At
  very low velocities you could also 
  expect van der Waals forces between the pearls causing them to scatter.
  We hope in a later article to be able to estimate the long distance and
  therefore low velocity behaviour better.

\begin{figure}  
	\includegraphics{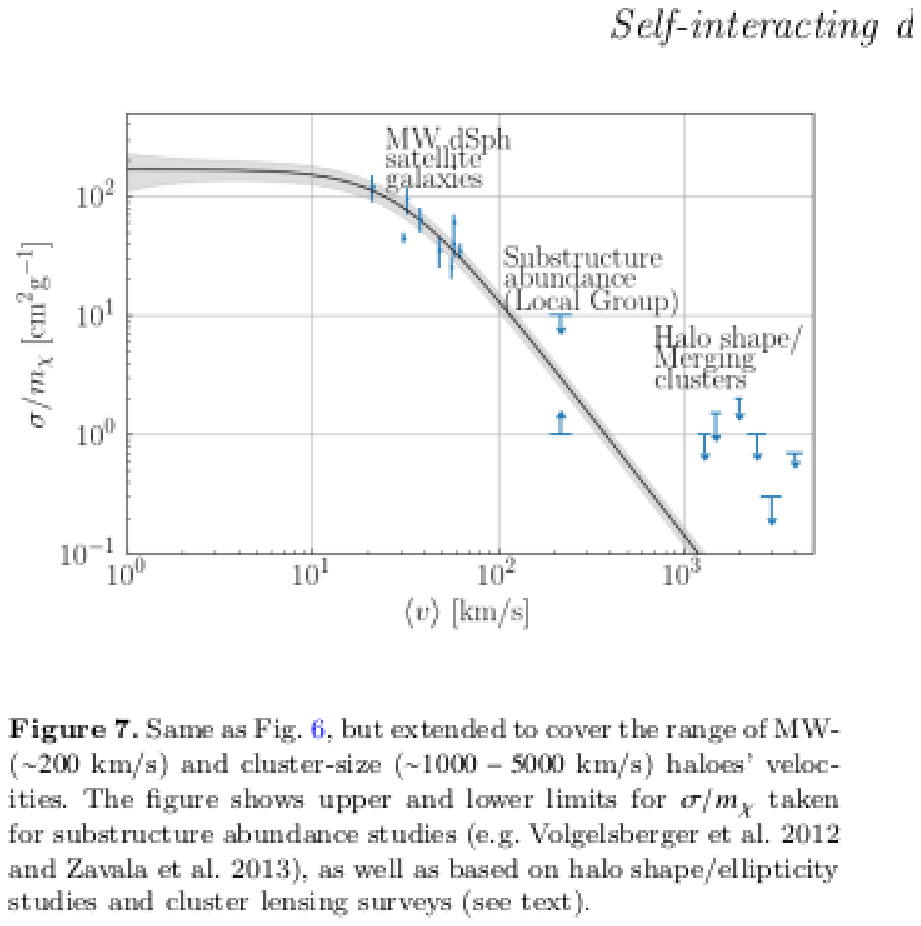}
	\caption{\label{Correa2} Cross section per unit mass $\frac{\sigma}{M}$ 
		of self
		interacting dark matter particles as a function of the collision velocity
		$v$ in dwarf galaxies from reference \cite{CAC}.}
\end{figure}

  In fact we find on Figure \ref{Correa2} from \cite{CAC}
that at the velocity $v=410 \, km/s$ the cross section to mass ratio
is very close to
\begin{eqnarray}
  \frac{\sigma}{M}|_{at \; v=410 km/s} &=& 1 \, cm^2/g 
\end{eqnarray}
 where the kinetic energy per cloud electron is 3.5 keV.
Thus taking it that the cross section to mass ratio we have calculated is indeed
rather that at 410 km/s, we would fit the mass to be 150 times bigger than the
value $7.0*10^5m_N $ from (\ref{corrMmin}). That would mean we have the mass
prediction
\begin{eqnarray}
  M&=& 1.0*10^{8}m_N \quad \hbox{ (for $\frac{\sigma}{M}=1 \, cm^2/g$ )} \label{M1}
\end{eqnarray}  
 corresponding to a pearl radius\footnote{We note that the increase in the
 value of $R$ compared to (\ref{corrRmin}) leads to a similar increase in
 $r_{cloud \, 3.5 keV}$ and consequently to a value of 
 $\frac{\sigma}{M} \sim 2 \, cm^2/g$.}
\begin{eqnarray}
R&=& 4.3*10^{-11}m.\label{R1} 
\end{eqnarray}

However the mass (\ref{M1}) is still much smaller than the minimal mass
$M_{minimal}$ (\ref{minMyear}) needed to avoid wash out of the DAMA 
seasonal effect.

 \subsection{Improvement by Dirt Surrounding the Pearl?}
 \label{dirtypearl}

Taking it seriously that we at least have some observations
suggesting a dark matter self interaction, having  
a ratio $\frac{\sigma}{M}$ which approaches e.g. the value $150 \, cm^2/g$, we
unavoidably need that the dark matter particle has some extension up to
a cross section that can be consistent with such a ratio. But now
the ratio {\em falls} as a function of the mass $M$ or the radius,
whether $R$ or $r_{cloud \; 3.5 keV}$ is used. 
So if we make the mass and the radius as big as required by (\ref{minMyear}) 
and (\ref{minRyear}), i.e. $M \ge 1.2*10^{12} \, GeV$ and 
$R \ge 0.99*10^{-9}m$, 
we cannot match the ratio $\frac{\sigma}{M} = 150 \, cm^2/g$ without changing
our model. The idea we propose is to assume that during their passage
through space - probably at a rather early stage when there was a lot of
ordinary matter around - the pearls have collected some dirt
around them. They could have become small grains of dust having used the
original pearl as a seed. Such dust would again be washed off / shaken off
in the Earth's atmosphere or the shielding mountains of the laboratories.
If so the dust would not influence the passage through the earth material.

Now for a pearl of mass $M=1.2*10^{12} \, GeV$ and radius $R=0.99*10^{-9} \, m$ 
the ratio $\frac{\sigma}{M}=1.5*10^{-2} \, cm^2/g$. 
So in order to increase this ratio to
$150 \, cm^2/g$ we need the speculated dirt to increase the cross section
of the dark matter particle by a factor of $10^4$.  
But now the
cross section of course goes as the square of the radius and thus the dirt
must increase the radius by a factor 100.

\begin{figure}
	\includegraphics{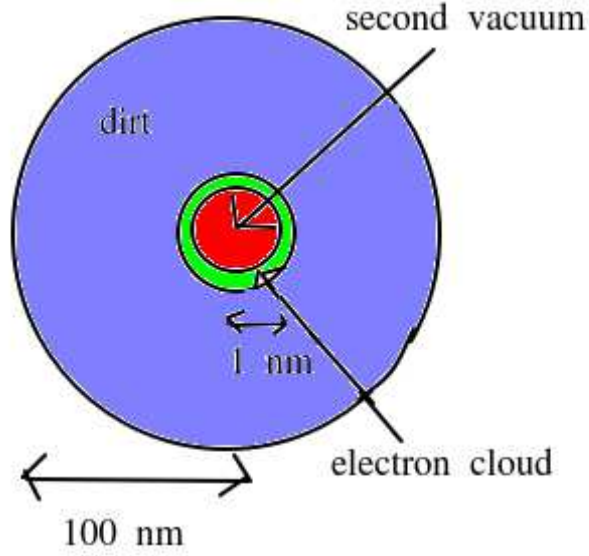} 
	\caption{\label{dirt} The blueish sphere around the pearl symbolizes 
		the ``dirt'', which may be attached to the pearl and is likely 
		to be some form of carbon 
		or carbohydrate or silicate.
		This dirt would be stripped off when the pearl passes through 
		the atmosphere or the Earth. 
		The inner red sphere is the second vacuum part, and in between one sees a relatively tiny region comprising the electron cloud.} 
\end{figure}

On Figure \ref{dirt} we have indicated how our pearl 
now with dimension
$R=0.99*10^{-9} \, m$ or bigger is enhanced by an 100 times bigger in radius
dust configuration around this genuine pearl. Since the genuine pearl is
about $1 \, nm$ in radius the full pearl radius becomes of the order of $100 \, nm$.

\section{\label{sec5} Numerical Rates for DAMA and
    Xenon1T-electron-recoil-excess}
      \subsection{The Kinetic Energy Flux from Dark Matter}
      \label{kef}
      The dark matter density $D_{sol}$ in our part of the Milky Way and its
      velocity $v$ are of the orders of magnitude
\begin{eqnarray}
D_{sol} &=& 0.3 \; GeV/cm^3\\
&=& 5.34 *10^{-22}kg/m^3\\
v&=& 300 \; km/s \quad \hbox{(relative to solar system)}.
\end{eqnarray}
This gives a kinetic energy density
\begin{eqnarray}
D_{kin \; energy} &=& \frac{1}{2}v^2 D_{sol}\\
&=& 0.5*(10^{-3})^2c^2* 5.34*10^{-22}kg/m^3\\
&=&  
2.40 *10^{-11}J/m^3,
\end{eqnarray}
meaning an influx of kinetic energy
\begin{eqnarray}
\hbox{``power per $m^2$''}&=& vD_{kin \; energy}\\
&=& 7.2*10^{-6}W/m^2.
\end{eqnarray}

Distributing this energy rate over the amount of matter down to
the depth $1400 \; m$ with density $3000 \;kg/m^3$ we obtain the energy rate
per kg

\begin{eqnarray}
  \hbox{`` power to deposit''}&=& \frac{7.2*10^{-6}W/m^2}{1400 \; m * 3000 \; kg/m^3}\\
  &=& 1.7*10^{-12}W/kg.
  \end{eqnarray}

However, assuming that all the events from the
dark matter - as given by the modulated part of the signal found by
DAMA-LIBRA -
are just due to decays with the decay energy 3.5 keV, the rate of energy
deposition per kg observed by DAMA-LIBRA \cite{DAMA2} is
\begin{eqnarray}
  \hbox{``deposited rate ''} &=& 0.0412 \; cpd/kg*3.5 \; keV\\
  &=& \frac{0.0412 \; cpd/kg*3.5*1.6*10^{-16}J }{86400 \; s/day}\\
 &=& 2.7*10^{-22}W/kg,
 \end{eqnarray}
  which is
\begin{eqnarray}
\frac{2.7*10^{-22}W/m^2}{1.7*10^{-12}W/m^2}
  &=& 1.6 *10^{-10} \; \hbox{times as much.}
  \end{eqnarray}

We can express this by saying that there is a need for a suppression factor
$suppression$ being $1.6*10^{-10}$ for the DAMA-LIBRA rate. For the excess
of the electronic recoil events found by Xenon1T the corresponding suppression
factor must be the 250 times smaller number. This is because the event rate
of these
excess electron recoil events is 250 times smaller than 
that of the annual modulation
part of the DAMA rate and the depth of the experiment under the earth is the
same 1400 m. Thus we summarize the {\em experimentally} determined suppression
factors:
\begin{eqnarray}
  suppression_{DAMA} &=& 1.6*10^{-10}\\
  suppression_{Xenon1T}&=& \frac{1.6*10^{-10}}{250}=  6.4*10^{-13}.
\end{eqnarray}
\subsection{Estimating ``suppression'' theoretically}

The idea for obtaining theoretical estimates of these suppression factors
is to say that the observed events come from excitations of our pearls with a
lifetime of the order of the time it takes for the pearl, after its excitation
under its stopping in the air or in the stone above the experiments, to reach
down to the experimental detectors.

As discussed in section \ref{discuss}, there is a tension between the 
high mass of the pearl required for it to reach through 1400 m of earth 
to the detectors in less than a year and the value of 
$\frac{\sigma}{M} = 150 \, cm^2/g$ suggested by the dwarf galaxy observations.
Thus we are driven to seek a passage time for the pearl through the earth 
material close to the upper limit of one year from the condition that 
the seasonal variation should not wash away.

\subsection{Equally hard to excite and to de-excite}

In order that there can be any de-excitations of the pearls after such
a year,
it is of course needed that an appreciable part of the possible 3.5 keV
excitations
of our pearls have lifetimes of this order of magnitude.
A priori these excitations are excitons for which the electron and hole can
be close by and decay rapidly or it is possible that one of the partners is
outside in the electron cloud and long lived.
By arguing that some tunnelling of electrons in or out or around in the
pearl may be needed for some (de-)excitations, we can claim that
the lifetimes for the various excitation possibilities are smoothly distributed
over a wide range in the logarithm of the lifetime;
then there will be some pearl-excitations with the appropriate
lifetime, although somewhat suppressed by a factor of the order of
$1/width$ where the $width$ here is the width of the logarithmic distribution.
We shall take this $width$ to be of order $\ln (suppression_{DAMA}) \sim 23$. But
more importantly:

If a certain excitation is long-lived, it is also hard to produce.
So we shall talk about an effective `` stopping'' or ``filling time''
for a pearl passing into the Earth, and imagine that during this
``stopping'' or ``filling time'' the excitations of the pearls
have to be created. So the probability for excitation or
$suppression$ would be expected to be
\begin{eqnarray}
  suppression &\approx& \frac{\hbox{``filling time''}}{\hbox{``lifetime''}}.
\end{eqnarray}
   If the excitation happens to be
   of sufficiently long lifetime - say of order
   one year - then we can expect it
to have a sensible chance of de-exciting just in the experimental
detectors in Gran Sasso, DAMA or Xenon1T say.

But what shall we take for this ``stopping'' or ``filling time''?
A relatively simple idea, which is presumably right, is to say that
the stopping takes place high in the atmosphere
because a pearl entering the Earth's atmosphere with galactic speed
will be slowed down
in the high air with a $\frac{\sigma}{M} \sim 2 \; cm^2/g$.
Now the density of the atmosphere rises by a factor $e=2.718...$
per about 7 km. So as the slowing down begins it will, because of this rising
density, essentially stop again after 7 km. Thus the time during which
the pearl is truly slowing down in speed and forming 3.5 keV excitations
is of the order of the time
it takes for it to run 7 km. With the pearl velocity of about $300 \; km/s$
(essentially the
escape velocity for the galaxy) we then have
\begin{eqnarray}
  \hbox{``stopping time''}&\approx& \frac{7 \; km}{ 300 \; km/s}\\
  &=& 0.023 \; s
    \end{eqnarray}

The crucial factor, which we believe to be most important, is that in
order to excite an excitation with a lifetime of the order
$3*10^7s=1 \, year$
it would a priori need
$3*10^7 s$ so that, if we only have 0.023 s, then
there will be a suppression:
\begin{eqnarray}
   suppression &=&
  \frac{\hbox{``stopping time''}}{\hbox{``lifetime''}}\\
    &\approx&  \frac{\hbox{``stopping time''}}{\hbox{``passage time''}}\\
  &\approx& \frac{0.023 \;s}{
  3*10^7 s}\\
  &=&
  6*10^{-10}
\end{eqnarray}

This crudest estimate has to be compared with the experimental suppressions
given above
\begin{eqnarray}
  suppression_{DAMA} &=& 1.6*10^{-10}\\
  \Rightarrow \frac{suppression_{theory}}{suppression_{DAMA}}&=&
  \frac{
  6*10^{-10}}
              {1.6*10^{-10}} \\
              &=&
              4\\
              suppression_{Xenon1T}&=& \frac{1.6*10^{-10}}{250}=  6.4*10^{-13}\\
              \Rightarrow \frac{suppression_{theory}}{suppression_{Xenon1T}}
              &=& \frac{
              6*10^{-10}}{6.4*10^{-13}}\\
              &\approx&
              1000
\end{eqnarray}

But the last number 1000 for the misfitting of the Xenon1T-excess should be
corrected for the fact that the time spent in the liquid xenon is shorter 
than in the solids, in which the pearls move slower.  

But here in addition there can be several corrections to $suppression_{theory}$, 
at least we should correct by  the
width in logarithm of the supposed distribution of the lifetimes  among the
different excitations. Above we suggested a factor 23, which would bring the
DAMA rate to only deviate by about a factor
6, now to the side that the theoretical suppression suppresses a factor 6
more than the experimental.  If it happens as we hope
that the pearl's speed through the liquid xenon is just 250 times
faster than through the solids, the deviation for the Xenon1T excess
will be just the same as for DAMA. Our estimate is of course
extremely uncertain.

We can never get the DAMA rate and the electron recoil excess rate from
Xenon1T agree
with the same estimate, in as far as they deviate by a factor 250,
unless we have some mechanism like the faster passage through the xenon
fluid because of its lower viscosity. 
Our only
chance is in a later paper to justify say the story that, because the
scintillator in which the Xenon1T events are observed is {\em fluid} while the
NaI in the DAMA experiment is solid, the pearls pass much faster through
the Xenon1T apparatus than they pass through the DAMA instrument. Imagine say
that the pearls partly hang and get stuck in the DAMA experiment, but that
they cannot avoid flowing down all the time while they are in the fluid xenon
in the Xenon1T scintillator.

\section{3.5 keV}
\label{sec7}
  { Order of magnitudewise we see 3.5 keV in {\huge 3} different places.}

  \includegraphics[scale=0.8]{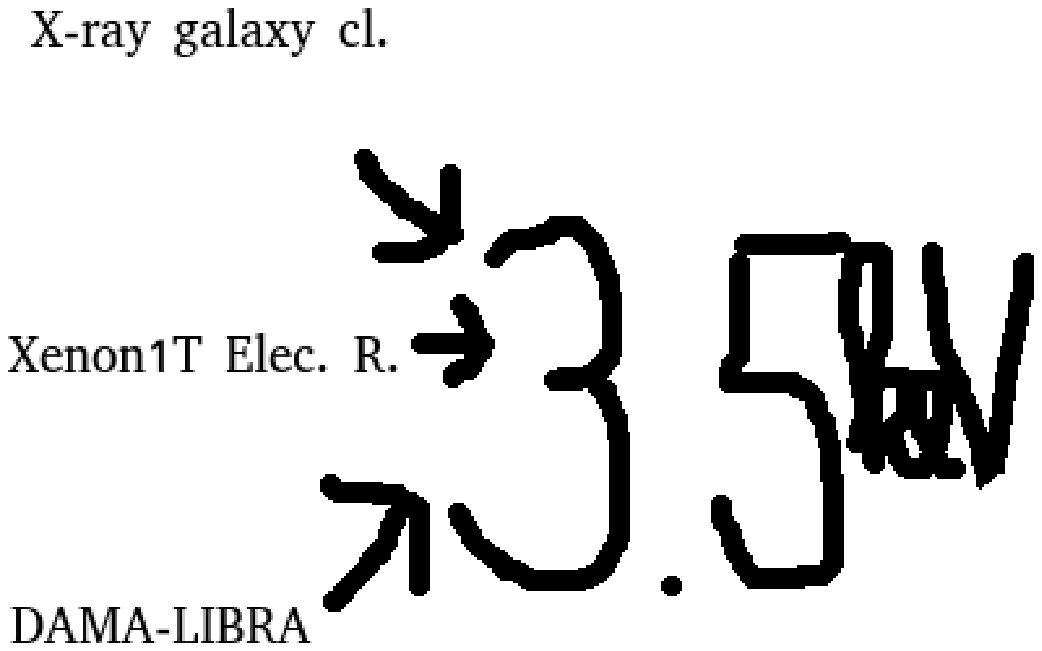}



  The energy level difference of about 3.5 keV occurring in 3 different
  places is important evidence motivating our
  model of dark matter particles being excitable by 3.5 keV:
  \begin{itemize}
  \item{\color{blue} The line} From places in outer space with a lot of dark
    matter, galaxy clusters, Andromeda and the Milky Way Center, an unexpected
    X-ray line with photon energies of 3.5 keV (to be corrected for Hubble
    expansion...) was seen.
  \item{\color{blue} Xenon1T} The dark matter experiment Xenon1T did not find
    the standard nucleus-recoil events expected for dark matter, 
    but found an excess of {\em electron-recoil}
    events with energies concentrated crudely around 3.5 keV.
  \item{\color{blue} DAMA} The seasonally varying component of their events lie in energy
    between 2 keV and 6 keV, not far from centering around 3.5 keV.
    \end{itemize}



  We take it seriously and not as an accident that both DAMA and Xenon1T see
  events with energies of the order of the controversial astronomical 3.5 keV X-ray
  line. We are thereby driven towards the hypothesis that the energies
  for the events in these underground experiments are
  determined from a decay
  of an excited particle, rather than from a collision with a particle in the
  scintillator material. It would namely be a pure accident, if a collision
  energy should just coincide with the dark matter excitation energy
  observed astronomically.

  {\em So we ought to have decays rather than collisions!}
  {\em How then can the dark matter particles get excited ?}

  You can think of the dark matter pearls in our model hitting electrons
  and/or nuclei on their way into the shielding:
  \begin{itemize}
  \item{\color{blue} Electrons} Electrons moving with the speed of the dark
    matter of the order of 300 km/s toward the pearls in the pearl frame
    will have kinetic energy of the order
    \begin{eqnarray}
      E_e &\approx & \frac{1}{2}*0.5 \; MeV*(\frac{300 \; km/s}{3*10^5km/s})^2=
      0.25 \; eV.
    \end{eqnarray}

  \item{\color{blue}Nuclei} If the nuclei are say Si, the energy in the collision
    will be 28*1900 times larger $\sim 5*10^4$ * $0.25 \; eV$
    $\approx$ 10 keV. That would allow a 3.5 keV excitation.

    To deliver such $\approx$ 10 keV energy the nucleus should hit
    something harder than just an electron inside the pearls. It should
    preferably hit a nucleus, e.g. C, inside the pearl.
    \end{itemize}

  \section{Discussing Parameters}
  \label{parameters}

  We have considered pearls having a cloud of $Z = 5.3*10^4$ electrons
  as in \cite{Bled21} 
  and improved the calculation of the minimal size pearl, for 
  which we obtained the updated skin radius $R = r_{cloud \; 3.3 MeV}$ and 
  mass M:
  \begin{equation}
  	R = 8.2*10^{-12} \, m \quad \hbox{and} \quad M = 7.0*10^5 \, GeV 
  	\label{par1}	
  \end{equation}  
  The cross section to mass ratio for this minimal size pearl is
  \begin{equation}
  \frac{\sigma}{M} = \frac{\pi r_{cloud \; 3.5 keV}^2}{M} \simeq 150 \, cm^2/g
  \end{equation}
   in agreement with the low velocity limit estimated from the study of dwarf
   galaxies around the Milky Way. 
   
   We also considered the possibility that the cross section obtained by taking 
   $r_{cloud \; 3.5 keV}$ to be the outer radius of the electron cloud 
   corresponds to a collision in which the kinetic energy of the pearl per 
   cloud electron is equal to 3.5 keV. This occurs for a velocity 
   $ v \sim 410 \, km/s$ and, according to the dwarf galaxy fit of 
   \cite{CAC} in Figure \ref{Correa2}, corresponds to 
   $\frac{\sigma}{M} \sim 1 \, cm^2/g$ and hence to a pearl of mass 150 times 
   that of the minimal size pearl with the following parameters:
   \begin{equation}
   	R = 4.3* 10^{-11} \, m \quad \hbox{and} \quad M = 1.0* 10^8 \, GeV.
   	\label{par2}
   \end{equation}  

   However neither of the pearls with parameters (\ref{par1}) or (\ref{par2}) 
   is heavy enough to pass through the earth to the DAMA-LIBRA detector
   in less than one year. In fact one needs a mass of at least
   $M_{minimal} =1.2*10^{12} \, GeV$ in order for the pearl to reach down
   in less then one year and so we take as our favoured pearl parameters:
   \begin{equation}
   R = 1.0 * 10^{-9} \, m \quad \hbox{and} \quad M = 1.2* 10^{12} \, GeV.
   \end{equation} 
   This $1 \, nm$ size pearl has a cross section per mass ratio of 
   $\frac{\sigma}{M} \simeq 1.5* 10^{-2} \, cm^2/g $, but we would like this ratio to approach $150 \, cm^2/g$ as suggested by the dwarf galaxy data. 
   So we have speculated that these pearls collect dirt around them, as 
   illustrated in Figure \ref{dirt}, which is stripped off them when 
   they pass through the atmosphere or the Earth. This dirt increases the radius
   by two orders of magnitude, so that the full radius becomes of order 
   $100 \, nm$.
   

The pressure from the surface tension $S$ of the pearl skin is balanced by
the degeneracy pressure from the electrons with Fermi momentum $p_f$:
\begin{equation}
	\frac{2S}{R} = \frac{p_f^4}{12\pi^2}.
\end{equation}
Our one parameter fit to the intensity and frequency of the 3.5 keV line 
from galaxy clusters and the Milky Way Center 
\cite{Corfu2019, theline, Bled20} determines the Fermi momentum to be 
$p_f = 3.3 MeV$. Hence the surface tension is given by
\begin{eqnarray}
	S &=& \frac{(3.3 \, MeV)^4}{24\pi^2 R}\\
	&=& 2.6*10^{12} \, MeV^3/m* R.
\end{eqnarray}
So for the minimal size pearl with $R = 8.2* 10^{-12} \, m$ the surface tension
is
\begin{equation}
	S^{1/3} = 2.8 \, MeV
\end{equation}
and for the 1 nm size pearl
\begin{equation}
	S^{1/3} = 14 \, MeV.
\end{equation}

  \section{Conclusion}
\label{Conclusion}


  We have described a seemingly viable model for dark
    matter consisting of atomic size but macroscopic pearls.
    These pearls consist of a
    bubble of a new speculated type of vacuum containing some normal
    material - presumably carbon - under the high pressure of the skin
    (surface tension). The electrons in a pearl are partly pushed out 
    of the genuine bubble of the new vacuum phase.
    In order to accommodate the annual modulation of the DAMA-LIBRA data
    we now favour a heavier pearl than the minimal size pearl considered 
    in the Bled proceedings \cite{Bled21}.
    These heavier pearls each contain about
    $10^{12}$ nucleons 
    in the
    bubble of radius about
    $R=1.0*10^{-9}m$ and have a surface tension $S^{1/3} = 14$ MeV.
    Unlike the minimal size pearls they have a much smaller cross section
    to mass ratio than the value $\frac{\sigma}{M} \simeq 150 \, cm^2/g$ 
    suggested by the dwarf galaxy data. So we were led to speculate that 
    this value could be achieved if ``dirt" had collected around them
    (see Figure \ref{dirt}) 
    increasing the radius from about 1 nm to about 100 nm.
     




  We have compared the model or attempted to fit:

  \begin{itemize}

  \item { Astronomical suggestions for the self interaction of dark matter
    in addition to pure gravity.}

  \item {The astronomical 3.5 keV X-ray emission line found by satellites,
    supposedly from dark
    matter.}

  \item {The underground dark matter searches.}

    \end{itemize}


  { We list below the quantities we have crudely estimated:}
  \begin{enumerate}
  \item The low velocity cross section divided by mass.
  \item A priori we predict that the event rate per kg seen by DAMA-LIBRA 
  and the electron excess event rate per kg at Xenon1T should be the same;
  they are both supposed to come from electrons emitted under the 
  de-excitation of our pearls at the same depth.
  The observed ratio of rates is 250, which we hope to explain as due to
  our pearls falling appreciably faster through the fluid xenon than through
  the solid NaI and hence spending a much shorter time in the Xenon1T 
  detector than in the DAMA-LIBRA detector.
  
  \item The absolute rate of the two underground experiments.
    (But unfortunately unless we explain the ratio of the rates for the
    two experiments as say due to the different velocities through the
    scintillator materials,
    we cannot  of course  predict the absolute
    rate to be better than deviating by about a factor of 250 from 
    at least one of them.)
  \item The rate of emission of the 3.5 keV X-ray line from the Tycho
    supernova remnant \cite{Jeltema} due to the excitation of our 
    pearls by cosmic
    rays \cite{Bled20}.
  \item Relation between the frequency 3.5 keV and the overall emission
    rate of this X-ray line observed from galaxy clusters etc.
  \item We also previously predicted the ratio of dark matter to
    atomic matter (=``usual'' matter) in the Universe to be of order 5
    by consideration of the binding energies per nucleon in helium and heavier
    nuclei,
    assuming that the atomic matter at some
    time about
    1 s after the Big Bang was spit out from the pearls under a fusion
    explosion
    from He fusing into say C \cite{Dark1}.
   \end{enumerate}

  \subsection{Outlook}

  At the end we want to mention a few ideas which we hope will be developed
  as a continuation of the present model:
\begin{itemize}
\item{\bf Speculative Phase from QCD.}
  QCD and even more QCD with  Nambu-Jona-Lasinio type spontaneous symmetry
  breaking
  is sufficiently complicated, that possibly a new phase appropriate for
  our pearls could be hiding there.
  There is already an extremely interesting observation \cite{KKS}.

\item{\bf Relative Rates of DAMA and Xenon1T.} A crucial test for our model is
  to reproduce the relative event rates in DAMA and in the excess
  of electron recoils in Xenon1T. This requires a careful study of the viscosity
  of fluid xenon and the properties of our pearls.

\item{\bf Walls in the Cosmos.} 
	If the domain walls have a sufficiently large surface energy density or 
	surface tension say $S^{1/3} \gtrsim 10$ MeV, their effect on 
	cosmology would be phenomenologically unacceptable. The pearls in this paper 
	have a surprisingly small surface tension ranging from $S^{1/3}$ = 2.8 
	MeV to 14 MeV. So astronomically extended domain walls just barely become 
	possible in our model e.g. walls around the
	large voids between the bands of galaxies; so that these voids could be say
	formally huge dark matter pearls, though with much smaller density.


\item{\bf New Experiments?} According to our estimates the observed rate of
  decays of our dark matter pearls should be larger the less shielding
  they pass through. So an obvious test of our model would be to make a
  DAMA-like
  experiment closer to the earth surface where we would expect a larger
  absolute rate than in DAMA, although there might of course be more background.
  Actually such an experiment is already being performed by the ANAIS group
  \cite{ANAIS},
  but they have so far failed to see an annual modulation in their event rate.

\end{itemize}

\section*{Acknowledgement}
HBN thanks the Niels Bohr Institute for his stay there as emeritus.
CDF thanks Glasgow University and the Niels Bohr Institute for hospitality and support.
Also we want to thank many colleagues for discussions and for organizing
conferences, where we have discussed previous versions of the present model,
especially the Bled and Corfu meetings.

We want to thank Konstantin Zioutas for calling our attention to 
earlier work on perhaps (in a hidden way) a similar model to ours, for very
interesting email correspondence and for inviting the present extended
version of our 2021 Bled-proceedings article.

\end{document}